\expandafter \def \csname CHAPLABELintro\endcsname {1}
\expandafter \def \csname CHAPLABELchains\endcsname {2}
\expandafter \def \csname EQLABELkdef\endcsname {2.1?}
\expandafter \def \csname EQLABELhiggsch\endcsname {2.2?}
\expandafter \def \csname EQLABELvts\endcsname {2.3?}
\expandafter \def \csname EQLABELanom\endcsname {2.4?}
\expandafter \def \csname EQLABELini\endcsname {2.5?}
\expandafter \def \csname EQLABELn2s\endcsname {2.6?}
\expandafter \def \csname EQLABELn1s\endcsname {2.7?}
\expandafter \def \csname EQLABELnhs\endcsname {2.8?}
\expandafter \def \csname EQLABELhter\endcsname {2.9?}
\expandafter \def \csname EQLABELh2\endcsname {2.10?}
\expandafter \def \csname EQLABELh22\endcsname {2.11?}
\expandafter \def \csname EQLABELenhG\endcsname {2.12?}
\expandafter \def \csname TABLABELkchain\endcsname {2.1?}
\expandafter \def \csname EQLABELnewinst\endcsname {2.13?}
\expandafter \def \csname EQLABELnewanom\endcsname {2.14?}
\expandafter \def \csname EQLABELhtern\endcsname {2.15?}
\expandafter \def \csname EQLABELhodgeformulas\endcsname {2.16?}
\expandafter \def \csname TABLABELterminal\endcsname {2.2?}
\expandafter \def \csname CHAPLABELpolyhedra\endcsname {3}
\expandafter \def \csname TABLABELchainone\endcsname {3.1?}
\expandafter \def \csname EQLABELsequ\endcsname {3.1?}
\expandafter \def \csname FIGLABELnabla\endcsname {3.1?}
\expandafter \def \csname EQLABELnk3\endcsname {3.2?}
\expandafter \def \csname EQLABELnose\endcsname {3.3?}
\expandafter \def \csname FIGLABELtorus\endcsname {3.2?}
\expandafter \def \csname TABLABELbottoms\endcsname {3.2?}
\expandafter \def \csname TABLABELbottomstwo\endcsname {3.3?}
\expandafter \def \csname EQLABELsur\endcsname {3.4?}
\expandafter \def \csname TABLABELscaling\endcsname {3.4?}
\expandafter \def \csname TABLABELheads\endcsname {3.5?}
\expandafter \def \csname EQLABELhnp\endcsname {3.5?}
\expandafter \def \csname EQLABELhnp2\endcsname {3.6?}
\expandafter \def \csname EQLABELhodgeformulasnp\endcsname {3.7?}
\expandafter \def \csname EQLABELbatyrev\endcsname {3.8?}
\expandafter \def \csname FIGLABELarrowsa\endcsname {3.3?}
\expandafter \def \csname FIGLABELarrowsb\endcsname {3.4?}
\expandafter \def \csname CHAPLABELfin\endcsname {4}
\expandafter \def \csname CHAPLABELappendix\endcsname {-2}
\expandafter \def \csname TABLABELHodge\endcsname {-2.1?}
\expandafter \def \csname TABLABELmoreHodge\endcsname {-2.2?}
\expandafter \def \csname FIGLABELfiga\endcsname {-2.1?}
\expandafter \def \csname FIGLABELfigb\endcsname {-2.2?}
\expandafter \def \csname FIGLABELfigc\endcsname {-2.3?}
\expandafter \def \csname FIGLABELfigd\endcsname {-2.4?}
\expandafter \def \csname FIGLABELfige\endcsname {-2.5?}
\expandafter \def \csname FIGLABELfigf\endcsname {-2.6?}
\expandafter \def \csname FIGLABELfigg\endcsname {-2.7?}
\expandafter \def \csname FIGLABELfigh\endcsname {-2.8?}
\expandafter \def \csname FIGLABELfiga\endcsname {-2.9?}
\expandafter \def \csname FIGLABELfigb\endcsname {-2.10?}
\expandafter \def \csname FIGLABELfigc\endcsname {-2.11?}
\expandafter \def \csname FIGLABELfigd\endcsname {-2.12?}
\expandafter \def \csname FIGLABELfige\endcsname {-2.13?}
\expandafter \def \csname FIGLABELfigf\endcsname {-2.14?}

\font\eightrm=cmr8 at 8pt

\font\seventeenrm=cmr17 at 17pt
\font\twentyonerm=cmr17 at 21pt

\font\ss=cmss10

\font\csc=cmcsc10

\font\twelvecal=cmsy10 at 12pt

\font\twelvemath=cmmi12

\font\seventeenbold=cmbx7 at 17pt

\font\fively=lasy5
\font\sevenly=lasy7
\font\tenly=lasy10

\textfont10=\tenly
\scriptfont10=\sevenly    
\scriptscriptfont10=\fively
\magnification=1200
\parskip=10pt
\parindent=20pt
\def\today{\ifcase\month\or January\or February\or March\or April\or May\or June
       \or July\or August\or September\or October\or November\or December\fi
       \space\number\day, \number\year}

\def\title#1{\footline={\ifnum\pageno<2\hfil
       \else\hss\tenrm\folio\hss\fi}\vskip1truein\centerline{{#1}   
       \footnote{\raise1ex\hbox{*}}{\eightrm Supported in part
       by the Robert A. Welch Foundation and N.S.F. Grants 
       PHY-880637 and\break PHY-8605978.}}}

\def\newpage{\vfill\eject}
\def\abstract#1{\centerline{\bf ABSTRACT}\vskip.2truein{\narrower\noindent#1
       \smallskip}}
\def\acknowledgements{\noindent\line{\bf Acknowledgements\hfill}\nobreak
    \vskip.1truein\nobreak\noindent\ignorespaces}
\def\runninghead#1#2{\voffset=2\baselineskip\nopagenumbers
       \headline={\ifodd\pageno\rightheadline\else \leftheadline\fi}
       \def\rightheadline{{\sl#1}\hfill{\rm\folio}}
       \def\leftheadline{{\rm\folio}\hfill{\sl#2}}}

\newcount\footnoteno
\def\Footnote#1{\advance\footnoteno by 1
                \let\SF=\empty 
                \ifhmode\edef\SF{\spacefactor=\the\spacefactor}\/\fi
                $^{\the\footnoteno}$\ignorespaces
                \SF\vfootnote{$^{\the\footnoteno}$}{#1}}

\def\figbox#1#2#3{\vbox{\vskip15pt
                   \vbox{\hrule
                    \hbox{\vrule
                     \vbox{\vskip12truept\centerline #1 \vskip6truept
                          {\hskip.4truein\vbox{\hsize=5truein\noindent
                          {\bf Figure\hskip5truept#2:}\hskip5truept#3}}
                     \vskip18truept}
                    \vrule}
                   \hrule}}}
\def\place#1#2#3{\vbox to0pt{\kern-\parskip\kern-7pt
                             \kern-#2truein\hbox{\kern#1truein #3}
                             \vss}\nointerlineskip}
\def\figurecaption#1#2{\kern.75truein\vbox{\hsize=5truein\noindent{\bf Figure
    \figlabel{#1}:} #2}}
\def\tablecaption#1#2{\kern.75truein\lower12truept\hbox{\vbox{\hsize=5truein
    \noindent{\bf Table\hskip5truept\tablabel{#1}:} #2}}}
\def\boxed#1{\lower3pt\hbox{
                       \vbox{\hrule\hbox{\vrule
                         \vbox{\kern2pt\hbox{\kern3pt#1\kern3pt}\kern3pt}\vrule}
                         \hrule}}}
\def\a{\alpha}

\def\g{\gamma}
\def\d{\delta}\def\D{\Delta}

\def\th{\theta}

\def\l{\lambda}
\def\m{\mu}
\def\n{\nu}

\def\ca#1{\relax\ifmmode {{\cal #1}}\else $\cal #1$\fi}

\def\calb{{\cal B}}

\def\calm{{\cal M}}

\def\inbar{\vrule height1.5ex width.4pt depth0pt}
\def\IB{\relax{\rm I\kern-.18em B}}
\def\IC{\relax\hbox{\kern.25em$\inbar\kern-.3em{\rm C}$}}
\def\ID{\relax{\rm I\kern-.18em D}}
\def\IE{\relax{\rm I\kern-.18em E}}
\def\IF{\relax{\rm I\kern-.18em F}}
\def\IG{\relax\hbox{\kern.25em$\inbar\kern-.3em{\rm G}$}}
\def\IH{\relax{\rm I\kern-.18em H}}
\def\II{\relax{\rm I\kern-.18em I}}
\def\IK{\relax{\rm I\kern-.18em K}}
\def\IL{\relax{\rm I\kern-.18em L}}
\def\IM{\relax{\rm I\kern-.18em M}}
\def\IN{\relax{\rm I\kern-.18em N}}
\def\IO{\relax\hbox{\kern.25em$\inbar\kern-.3em{\rm O}$}}
\def\IP{\relax{\rm I\kern-.18em P}}
\def\IQ{\relax\hbox{\kern.25em$\inbar\kern-.3em{\rm Q}$}}
\def\IR{\relax{\rm I\kern-.18em R}}
\def\IZ{\relax\ifmmode\hbox{\ss Z\kern-.4em Z}\else{\ss Z\kern-.4em Z}\fi}
\def\IGa{\relax{\rm I}\kern-.18em\Gamma}
\def\IPi{\relax{\rm I}\kern-.18em\Pi}
\def\ITh{\relax\hbox{\kern.25em$\inbar\kern-.3em\Theta$}}
\def\IOm{\relax\thinspace\inbar\kern1.95pt\inbar\kern-5.525pt\Omega}


\def\ie{{\it i.e.,\ \/}}

\def\noblackboxes{\overfullrule=0pt}
\def\define{\buildrel\rm def\over =}

\def\cy{Calabi--Yau} 
\def\cym{Calabi--Yau manifold}
\def\cys{Calabi--Yau manifolds}

\def\H#1#2{\relax\ifmmode {H^{#1#2}}\else $H^{#1 #2}$\fi}
\def\M{\relax\ifmmode{\calm}\else $\calm$\fi}

\def\Bigcheck{\lower3.8pt\hbox{\smash{\hbox{{\twentyonerm \v{}}}}}}
\def\bigboldcheck{\smash{\hbox{{\seventeenbold\v{}}}}}

\def\Bighat{\lower3.8pt\hbox{\smash{\hbox{{\twentyonerm \^{}}}}}}

\def\Msharp{\relax\ifmmode{\calm^\sharp}\else $\smash{\calm^\sharp}$\fi}
\def\Mflat{\relax\ifmmode{\calm^\flat}\else $\smash{\calm^\flat}$\fi}
\def\preMcheck{\kern2pt\hbox{\Bigcheck\kern-12pt{$\cal M$}}}
\def\Mcheck{\relax\ifmmode\preMcheck\else $\preMcheck$\fi}
\def\preMhat{\kern2pt\hbox{\Bighat\kern-12pt{$\cal M$}}}
\def\Mhat{\relax\ifmmode\preMhat\else $\preMhat$\fi}

\def\Bsharp{\relax\ifmmode{\calb^\sharp}\else $\calb^\sharp$\fi}
\def\Bflat{\relax\ifmmode{\calb^\flat}\else $\calb^\flat$ \fi}
\def\preBcheck{\hbox{\Bigcheck\kern-9pt{$\cal B$}}}
\def\Bcheck{\relax\ifmmode\preBcheck\else $\preBcheck$\fi}
\def\preBhat{\hbox{\Bighat\kern-9pt{$\cal B$}}}
\def\Bhat{\relax\ifmmode\preBhat\else $\preBhat$\fi}

\def\figBcheck{\kern3pt\hbox{\raise1pt\hbox{\bigboldcheck}\kern-11pt
    {\twelvecal B}}}
\def\figBsharp{{\twelvecal B}\raise5pt\hbox{$\twelvemath\sharp$}}
\def\figBflat{{\twelvecal B}\raise5pt\hbox{$\twelvemath\flat$}}

\def\gcheck{\hbox{\lower2.5pt\hbox{\Bigcheck}\kern-8pt$\g$}}
\def\lhat{\hbox{\raise.5pt\hbox{\Bighat}\kern-8pt$\l$}}

\def\Fcheck{\kern2pt\hbox{\raise1pt\hbox{\Bigcheck}\kern-10pt{$\cal F$}}}
\def\Fhat{\kern2pt\hbox{\raise1pt\hbox{\Bighat}\kern-10pt{$\cal F$}}}
 
\def\cp#1{\relax\ifmmode {\IP\kern-2pt{}_{#1}}\else $\IP\kern-2pt{}_{#1}$\fi}
\def\h#1#2{\relax\ifmmode {b_{#1#2}}\else $b_{#1#2}$\fi}

\def\half{{1\over 2}}
\def\tr{{\rm tr}}
\def\frac#1#2{{#1\over #2}}

\def\cone{\relax\thinspace\hbox{$<\kern-.8em{)}$}}
\mathchardef\mho"0A30

\def\asymp{\sim}
\def\-{\hphantom{-}}


\def\npb#1{Nucl.\ Phys.\ {\bf B#1}}

\def\plb#1{Phys. Lett. {\bf #1B}}


\def\picture #1 by #2 (#3){\vbox to #2{\hrule width #1 height 0pt depth 0pt
                                       \vfill\special{picture #3}}}
\def\scaledpicture #1 by #2 (#3 scaled #4){{\dimen0=#1 \dimen1=#2
           \divide\dimen0 by 1000 \multiply\dimen0 by #4
            \divide\dimen1 by 1000 \multiply\dimen1 by #4
            \picture \dimen0 by \dimen1 (#3 scaled #4)}}
\def\illustration #1 by #2 (#3){\vbox to #2{\hrule width #1 height 0pt depth 0pt
                                       \vfill\special{illustration #3}}}
\def\scaledillustration #1 by #2 (#3 scaled #4){{\dimen0=#1 \dimen1=#2
           \divide\dimen0 by 1000 \multiply\dimen0 by #4
            \divide\dimen1 by 1000 \multiply\dimen1 by #4
            \illustration \dimen0 by \dimen1 (#3 scaled #4)}}


\def\delaOssa{\nobreak\vskip1truein\hbox to\hsize
       {\hskip 4truein Xenia de la Ossa\hfill}}

\def\hoy{\number\day\space de \ifcase\month\or enero\or febrero\or marzo\or
       abril\or mayo\or junio\or julio\or agosto\or septiembre\or octubre\or
       noviembre\or diciembre\fi\space de \number\year}

\def\cropen#1{\crcr\noalign{\vskip #1}}

\newif\ifproofmode
\proofmodefalse

\newif\ifforwardreference
\forwardreferencefalse

\newif\ifchapternumbers
\chapternumbersfalse

\newif\ifcontinuousnumbering
\continuousnumberingfalse

\newif\iffigurechapternumbers
\figurechapternumbersfalse

\newif\ifcontinuousfigurenumbering
\continuousfigurenumberingfalse

\newif\iftablechapternumbers
\tablechapternumbersfalse

\newif\ifcontinuoustablenumbering
\continuoustablenumberingfalse

\font\eqsixrm=cmr6

\def\marginstyle{\eqsixrm}

\newtoks\chapletter
\newcount\chapno
\newcount\eqlabelno
\newcount\figureno
\newcount\tableno

\chapno=0
\eqlabelno=0
\figureno=0
\tableno=0

\def\chapfolio{\ifnum\chapno>0 \the\chapno\else\the\chapletter\fi}

\def\bumpchapno{\ifnum\chapno>-1 \global\advance\chapno by 1
\else\global\advance\chapno by -1 \setletter\chapno\fi
\ifcontinuousnumbering\else\global\eqlabelno=0 \fi
\ifcontinuousfigurenumbering\else\global\figureno=0 \fi
\ifcontinuoustablenumbering\else\global\tableno=0 \fi}

\def\setletter#1{\ifcase-#1{}\or{}%
\or\global\chapletter={A}%
\or\global\chapletter={B}%
\or\global\chapletter={C}%
\or\global\chapletter={D}%
\or\global\chapletter={E}%
\or\global\chapletter={F}%
\or\global\chapletter={G}%
\or\global\chapletter={H}%
\or\global\chapletter={I}%
\or\global\chapletter={J}%
\or\global\chapletter={K}%
\or\global\chapletter={L}%
\or\global\chapletter={M}%
\or\global\chapletter={N}%
\or\global\chapletter={O}%
\or\global\chapletter={P}%
\or\global\chapletter={Q}%
\or\global\chapletter={R}%
\or\global\chapletter={S}%
\or\global\chapletter={T}%
\or\global\chapletter={U}%
\or\global\chapletter={V}%
\or\global\chapletter={W}%
\or\global\chapletter={X}%
\or\global\chapletter={Y}%
\or\global\chapletter={Z}\fi}

\def\tempsetletter#1{\ifcase-#1{}\or{}%
\or\global\chapletter={A}%
\or\global\chapletter={B}%
\or\global\chapletter={C}%
\or\global\chapletter={D}%
\or\global\chapletter={E}%
\or\global\chapletter={F}%
\or\global\chapletter={G}%
\or\global\chapletter={H}%
\or\global\chapletter={I}%
\or\global\chapletter={J}%
\or\global\chapletter={K}%
\or\global\chapletter={L}%
\or\global\chapletter={M}%
\or\global\chapletter={N}%
\or\global\chapletter={O}%
\or\global\chapletter={P}%
\or\global\chapletter={Q}%
\or\global\chapletter={R}%
\or\global\chapletter={S}%
\or\global\chapletter={T}%
\or\global\chapletter={U}%
\or\global\chapletter={V}%
\or\global\chapletter={W}%
\or\global\chapletter={X}%
\or\global\chapletter={Y}%
\or\global\chapletter={Z}\fi}

\def\chapshow#1{\ifnum#1>0 \relax#1%
\else{\tempsetletter{\number#1}\chapno=#1\chapfolio}\fi}

\def\chaplabel#1{\bumpchapno\ifproofmode\ifforwardreference
\immediate\write1{\noexpand\expandafter\noexpand\def
\noexpand\csname CHAPLABEL#1\endcsname{\the\chapno}}\fi\fi
\global\expandafter\edef\csname CHAPLABEL#1\endcsname
{\the\chapno}\ifproofmode\llap{\hbox{\marginstyle #1\ }}\fi\chapfolio}

\def\eqnum{\global\advance\eqlabelno by 1
\eqno(\ifchapternumbers\chapfolio.\fi\the\eqlabelno)}

\def\eqlabel#1{\global\advance\eqlabelno by 1 \ifproofmode\ifforwardreference
\immediate\write1{\noexpand\expandafter\noexpand\def
\noexpand\csname EQLABEL#1\endcsname{\the\chapno.\the\eqlabelno?}}\fi\fi
\global\expandafter\edef\csname EQLABEL#1\endcsname
{\the\chapno.\the\eqlabelno?}\eqno(\ifchapternumbers\chapfolio.\fi
\the\eqlabelno)\ifproofmode\rlap{\hbox{\marginstyle #1}}\fi}

\def\eqalignnum{\global\advance\eqlabelno by 1
&(\ifchapternumbers\chapfolio.\fi\the\eqlabelno)}

\def\eqalignlabel#1{\global\advance\eqlabelno by 1 \ifproofmode 
\ifforwardreference\immediate\write1{\noexpand\expandafter\noexpand\def
\noexpand\csname EQLABEL#1\endcsname{\the\chapno.\the\eqlabelno?}}\fi\fi
\global\expandafter\edef\csname EQLABEL#1\endcsname
{\the\chapno.\the\eqlabelno?}&(\ifchapternumbers\chapfolio.\fi
\the\eqlabelno)\ifproofmode\rlap{\hbox{\marginstyle #1}}\fi}

\def\eqref#1{\hbox{(\ifundefined{EQLABEL#1}***)\ifproofmode\ifforwardreference%
\else\write16{ ***Undefined Equation Reference #1*** }\fi
\else\write16{ ***Undefined Equation Reference #1*** }\fi
\else\edef\LABxx{\getlabel{EQLABEL#1}}%
\def\LAByy{\expandafter\stripchap\LABxx}\ifchapternumbers%
\chapshow{\LAByy}.\expandafter\stripeq\LABxx%
\else\ifnum\number\LAByy=\chapno\relax\expandafter\stripeq\LABxx%
\else\chapshow{\LAByy}.\expandafter\stripeq\LABxx\fi\fi)\fi}%
\ifproofmode\write2{Equation #1}\fi}

\def\fignum{\global\advance\figureno by 1
\relax\iffigurechapternumbers\chapfolio.\fi\the\figureno}

\def\figlabel#1{\global\advance\figureno by 1
\relax\ifproofmode\ifforwardreference
\immediate\write1{\noexpand\expandafter\noexpand\def
\noexpand\csname FIGLABEL#1\endcsname{\the\chapno.\the\figureno?}}\fi\fi
\global\expandafter\edef\csname FIGLABEL#1\endcsname
{\the\chapno.\the\figureno?}\iffigurechapternumbers\chapfolio.\fi
\ifproofmode\llap{\hbox{\marginstyle#1
\kern1.2truein}}\relax\fi\the\figureno}

\def\figref#1{\hbox{\ifundefined{FIGLABEL#1}!!!!\ifproofmode\ifforwardreference%
\else\write16{ ***Undefined Figure Reference #1*** }\fi
\else\write16{ ***Undefined Figure Reference #1*** }\fi
\else\edef\LABxx{\getlabel{FIGLABEL#1}}%
\def\LAByy{\expandafter\stripchap\LABxx}\iffigurechapternumbers%
\chapshow{\LAByy}.\expandafter\stripeq\LABxx%
\else\ifnum \number\LAByy=\chapno\relax\expandafter\stripeq\LABxx%
\else\chapshow{\LAByy}.\expandafter\stripeq\LABxx\fi\fi\fi}%
\ifproofmode\write2{Figure #1}\fi}

\def\tabnum{\global\advance\tableno by 1
\relax\iftablechapternumbers\chapfolio.\fi\the\tableno}

\def\tablabel#1{\global\advance\tableno by 1
\relax\ifproofmode\ifforwardreference
\immediate\write1{\noexpand\expandafter\noexpand\def
\noexpand\csname TABLABEL#1\endcsname{\the\chapno.\the\tableno?}}\fi\fi
\global\expandafter\edef\csname TABLABEL#1\endcsname
{\the\chapno.\the\tableno?}\iftablechapternumbers\chapfolio.\fi
\ifproofmode\llap{\hbox{\marginstyle#1
\kern1.2truein}}\relax\fi\the\tableno}

\def\tabref#1{\hbox{\ifundefined{TABLABEL#1}!!!!\ifproofmode\ifforwardreference%
\else\write16{ ***Undefined Table Reference #1*** }\fi
\else\write16{ ***Undefined Table Reference #1*** }\fi
\else\edef\LABtt{\getlabel{TABLABEL#1}}%
\def\LABTT{\expandafter\stripchap\LABtt}\iftablechapternumbers%
\chapshow{\LABTT}.\expandafter\stripeq\LABtt%
\else\ifnum\number\LABTT=\chapno\relax\expandafter\stripeq\LABtt%
\else\chapshow{\LABTT}.\expandafter\stripeq\LABtt\fi\fi\fi}%
\ifproofmode\write2{Table#1}\fi}

\newdimen\sectionskip     \sectionskip=20truept
\newcount\sectno
\def\section#1#2{\sectno=0 \null\vskip\sectionskip
    \centerline{\chaplabel{#1}.~~{\bf#2}}\nobreak\vskip.2truein
    \noindent\ignorespaces}

\def\advancesectno{\global\advance\sectno by 1}
\def\sectfolio{\number\sectno}
\def\subsection#1{\goodbreak\advancesectno\null\vskip10pt
                  \noindent\chapfolio.~\sectfolio.~{\bf #1}
                  \nobreak\vskip.05truein\noindent\ignorespaces}

\def\uttg#1{\null\vskip.1truein
    \ifproofmode \line{\hfill{\bf Draft}:
    UTTG--{#1}--\number\year}\line{\hfill\today}
    \else \line{\hfill UTTG--{#1}--\number\year}
    \line{\hfill\ifcase\month\or January\or February\or March\or April\or May\or June
    \or July\or August\or September\or October\or November\or December\fi
    \space\number\year}\fi}

\def\contents{\noindent
   {\bf Contents\Z}\nobreak\vskip.05truein\noindent\ignorespaces}

\def\getlabel#1{\csname#1\endcsname}
\def\ifundefined#1{\expandafter\ifx\csname#1\endcsname\relax}
\def\stripchap#1.#2?{#1}
\def\stripeq#1.#2?{#2}

%
\catcode`@=11 
\def\space@ver#1{\let\@sf=\empty\ifmmode#1\else\ifhmode%
\edef\@sf{\spacefactor=\the\spacefactor}\unskip${}#1$\relax\fi\fi}
\newcount\referencecount     \referencecount=0
\newif\ifreferenceopen       \newwrite\referencewrite
\newtoks\rw@toks
\def\refmark#1{\relax[#1]}
\def\refend{\refmark{\number\referencecount}}
\newcount\lastrefsbegincount \lastrefsbegincount=0
\def\refsend{\refmark{\count255=\referencecount%
\advance\count255 by -\lastrefsbegincount%
\ifcase\count255 \number\referencecount%
\or\number\lastrefsbegincount,\number\referencecount%
\else\number\lastrefsbegincount-\number\referencecount\fi}}
\def\refch@ck{\chardef\rw@write=\referencewrite
\ifreferenceopen\else\referenceopentrue
\immediate\openout\referencewrite=referenc.texauxil \fi}
%
{\catcode`\^^M=\active 
  \gdef\obeyendofline{\catcode`\^^M\active \let^^M\ }}%
%
{\catcode`\^^M=\active 
  \gdef\ignoreendofline{\catcode`\^^M=5}}
{\obeyendofline\gdef\rw@start#1{\def\t@st{#1}\ifx\t@st\blankend%
\endgroup\@sf\relax\else\ifx\t@st\bl@nkend\endgroup\@sf\relax%
\else\rw@begin#1
\backtotext
\fi\fi}}
{\obeyendofline\gdef\rw@begin#1
{\def\n@xt{#1}\rw@toks={#1}\relax%
\rw@next}}
\def\blankend{}
{\obeylines\gdef\bl@nkend{
}}
\newif\iffirstrefline  \firstreflinetrue
\def\rwr@teswitch{\ifx\n@xt\blankend\let\n@xt=\rw@begin%
\else\iffirstrefline\global\firstreflinefalse%
\immediate\write\rw@write{\noexpand\obeyendofline\the\rw@toks}%
\let\n@xt=\rw@begin%
\else\ifx\n@xt\rw@@d \def\n@xt{\immediate\write\rw@write{%
\noexpand\ignoreendofline}\endgroup\@sf}%
\else\immediate\write\rw@write{\the\rw@toks}%
\let\n@xt=\rw@begin\fi\fi\fi}
\def\rw@next{\rwr@teswitch\n@xt}
\def\rw@@d{\backtotext} \let\rw@end=\relax
\let\backtotext=\relax

\newdimen\refindent     \refindent=30pt
\def\Textindent#1{\noindent\llap{#1\enspace}\ignorespaces}
\def\refitem#1{\par\hangafter=0 \hangindent=\refindent\Textindent{#1}}
\def\REFNUM#1{\space@ver{}\refch@ck\firstreflinetrue%
\global\advance\referencecount by 1 \xdef#1{\the\referencecount}}
\def\refnum#1{\space@ver{}\refch@ck\firstreflinetrue%
\global\advance\referencecount by 1\xdef#1{\the\referencecount}\refend}

\def\REF#1{\REFNUM#1%
\immediate\write\referencewrite{%
\noexpand\refitem{#1.}}%
\begingroup\obeyendofline\rw@start}
\def\ref{\refnum\?%
\immediate\write\referencewrite{\noexpand\refitem{\?.}}%
\begingroup\obeyendofline\rw@start}
\def\Ref#1{\refnum#1%
\immediate\write\referencewrite{\noexpand\refitem{#1.}}%
\begingroup\obeyendofline\rw@start}
\def\REFS#1{\REFNUM#1\global\lastrefsbegincount=\referencecount%
\immediate\write\referencewrite{\noexpand\refitem{#1.}}%
\begingroup\obeyendofline\rw@start}

\def\REFSCON#1{\REF#1}

\def\cite#1{\refmark#1}
\catcode`@=12 
%
\input epsf.tex
\proofmodefalse
\baselineskip=15pt plus 1pt minus 1pt
\parskip=5pt
\chapternumberstrue
\figurechapternumberstrue
\tablechapternumberstrue
\ifproofmode
\immediate\openout2=allcrossreferfile \fi
\ifforwardreference\input labelfile
\ifproofmode\immediate\openout1=labelfile \fi\fi
\noblackboxes
\hfuzz=1pt
\vfuzz=2pt


\def\hourandminute{\count255=\time\divide\count255 by 60
\xdef\hour{\number\count255}
\multiply\count255 by -60\advance\count255 by\time
\hour:\ifnum\count255<10 0\fi\the\count255}

\def\subsection#1{\goodbreak\advancesectno\null\vskip10pt
                  \noindent{\it \chapfolio.\sectfolio.~#1}
                  \nobreak\vskip.05truein\noindent\ignorespaces}
\def\cite#1{\refmark{#1}}
\def\\{\hfill\break}
\def\cropen#1{\crcr\noalign{\vskip #1}}
\def\contents{\line{{\bf Contents}\hfill}\nobreak\vskip.05truein\noindent%
              \ignorespaces}

\def\crm{\cr\noalign{\medskip}}
\def\crb{\cr\noalign{\bigskip}}
\def\point#1{\noindent\setbox0=\hbox{#1}\kern-\wd0\box0}

\def\tv{\tilde{v}}
\def\nab#1{{}^#1\nabla}
\nopagenumbers\pageno=0
\rightline{\eightrm UTTG-04-96}\vskip-5pt
\rightline{\eightrm hep-th/9603170}\vskip-5pt
\rightline{\eightrm 25 March 1996, revised 6 May 1996, 19 March 1997}

\vskip1truein
\centerline{\seventeenrm Duality Between the Webs of}
\vskip10pt
\centerline{\seventeenrm Heterotic and Type II Vacua}
\vskip.6truein
\centerline{\csc Philip~Candelas$^1$  and Anamar\'\i a~Font$^2$}
\vfootnote{$^{\eightrm 1}$}{\eightrm candelas@physics.utexas.edu.}
\vfootnote{$^{\eightrm 2}$}{\eightrm afont@dino.conicit.ve.
On sabbatical leave from Departamento de
F\'{\i}sica, Facultad de Ciencias,}
\vfootnote{}{\eightrm Universidad Central de Venezuela.}

\vskip.3truein\bigskip
\centerline{\it Theory Group}
\centerline{\it Department of Physics}
\centerline{\it University of Texas}
\centerline{\it Austin, TX 78712, USA}
\vskip1truein\bigskip
\nobreak\vbox{
\centerline{\bf ABSTRACT}
\vskip.25truein
\noindent We discuss how transitions in the space of heterotic $K3\times T^2$
compactifications are mapped by duality into transitions in the space of
Type II compactifications on \cys. We observe that perturbative
symmetry restoration, as well as non-perturbative processes such as
changes in the number of tensor multiplets, have at least in many cases a
simple description in terms of the reflexive polyhedra of the \cys.
Our results suggest that to many, perhaps all, four-dimensional
$N=2$ heterotic vacua there are corresponding type II
vacua.}
\newpage
{\baselineskip=13pt
\contents
\vskip15pt
\item{1.~}Introduction
\bigskip
\item{2.~}Heterotic Chains
\bigskip
\item{3.~}Sequences of Reflexive Polyhedra\medskip
\itemitem{3.1~}{\it General structure}\smallskip
\itemitem{3.2~}{\it $k=0$ and $k=$~half integer}\smallskip
\itemitem{3.3~}{\it Non-perturbative effects}\smallskip
\itemitem{3.4~}{\it Irreducibility}\smallskip
\itemitem{3.5~}{\it The final form of the polyhedra and the Dynkin diagrams}
\bigskip
\item{4.~}Discussion
\bigskip
\item{A.~}Appendix: Tables of Hodge Numbers and Figures
}
\newpage

\pageno=1
\headline={\ifproofmode\hfil\eightrm draft:\ \today\
\hourandminute\else\hfil\fi}
\footline={\rm\hfil\folio\hfil}
\section{intro}{Introduction}
In this paper we explore four-dimensional, $N\!=\!2$ string
vacua in connection with the conjectured duality~
\REFS\rKV{S.~Kachru and C.~Vafa, \npb{450} (1995) 69, hep-th/9505105.}
\REFSCON\rFHSV{
S.~Ferrara, J.~Harvey, A.~Strominger and C.~Vafa,\\
\plb{361} (1995) 59, hep-th/9505162.}
\refsend\
between $(0,4)$ compactifications of
the $E_8\times E_8$ heterotic string on the manifold $K3\times T^2$ and the
type IIA string compactified on a Calabi-Yau manifold.
This duality has been the subject of several articles in the recent
literature~
\REFS\rKLM{
A.~Klemm, W.~Lerche, and P.~Mayr, \plb{357} (1995) 313, hep-th/9506112.}
\REFSCON\rVW{
C.~Vafa and E.~Witten,  hep-th/9507050.}
\REFSCON\rVARI{
C.~G\'omez and E.~L\'opez,  \plb {356} (1995) 487, hep-th/9506024;\\
M.~Bill\'o, A.~Ceresole, R.~D'Auria, S.~Ferrara, P.~Fr\'e, T.~Regge,
P.~Soriani and A.~van~Proyen, hep-th/9506075;\\
I.~Antoniadis, E.~Gava, K.S.~Narain and T.R.~Taylor,\\
\npb{455} (1995) 109, hep-th/9507115;\\
G.~Lopes Cardoso, D.~Lust and T.~Mohaupt,\\
\npb{455} (1995) 131, hep-th/9507113;\\
G.~Curio, \plb{368} (1996) 78, hep-th/9509146.}
\REFSCON\rKLT{V.~Kaplunovsky, J.~Louis and S.~Theisen,\\
  \plb{357} (1995) 71, hep-th/9506110.}
\REFSCON\rKKLMV{S.~Kachru, A.~Klemm, W.~Lerche, P.~Mayr and C.~Vafa,\\
\npb{459}~(1996)~537, hep-th/9508155.}
\REFSCON\rAFIQ{ G.~Aldazabal, A.~Font, L.~E.~Ib\'a\~nez and F.~Quevedo,\\
\npb{461}~(1996)~85, hep-th/9510093.}
\REFSCON\rAL{P.~S.~Aspinwall and J. Louis, \plb{369} (1996) 233,
hep-th/9510234.}
\REFSCON\rAsp{P.~S.~Aspinwall, \plb{371} (1996) 231, hep-th/9511171.}
\REFSCON\rVARII{B.~Hunt and R.~Schimmrigk, hep-th/9512138;\\
R.~Blumenhagen and A.~Wisskirchen, hep-th/9601050.}
\refsend.
Schematically we may write
 $$
\hbox{Het}_{E_8\times E_8}\bigl[K3\times T^2,\,\ca{V}\bigr]\simeq
\hbox{IIA}\bigl[\ca{M}\bigr] $$
where $\ca{V}$ denotes the bundle (more properly a sheaf) corresponding to the
background gauge field and $\ca{M}=\ca{M}_\ca{V}$ a \cym\ that depends on
$\ca{V}$.  This duality induces a correspondence
 $$
\ca{V}\leftrightarrow \ca{M}_\ca{V}$$
between vector bundles on $K3\times T^2$ and certain (perhaps all) \cy\
manifolds, though a special role is played by manifolds that are
$K3$-fibrations \cite{\rKLM,\rVW,\rAL}.
The evidence supporting this correspondence is based on the identification of
certain dual pairs $(\ca{V},\ca{M}_\ca{V})$ through a computation of Hodge
numbers and, very compellingly, on detailed comparison of the structure of the
moduli-spaces of $\ca{V}$ and $\ca{M}_\ca{V}$ for candidate dual pairs
previously identified on the basis of their Hodge numbers
\REF\rKM{A.~Klemm and P.~Mayr hep-th/9601014.}
\REF\rBKK{P.~Berglund, S.~Katz and A.~Klemm, in preparation.}
\cite{\rKV,\rKLT,\rKKLMV,\rKM,\rBKK}.

Two observations concerning this duality are the subject of the present
article: that a great many \cys\ are known in terms of toric data
\REFS\rKleSch{A.~Klemm and R.~Schimmrigk, \npb{411} (1994) 559,
hep-th/9204060.}
\REFSCON\rKreSka{M.~Kreuzer and H.~Skarke, \npb{388}~(1992)~113,
hep-th/9205004.}
\REFSCON\rCDK{P.~Candelas, X. de la Ossa and S.~Katz,\\
\npb{450} (1995) 267, hep-th/9412117.}
\REFSCON\rSka{H.~Skarke, alg-geom/9603007.}
\refsend\
and the correspondence, pointed out by Batyrev, between \cys\ and reflexive
polyhedra~
\Ref\rBat{V.~Batyrev, Duke Math.\ Journ.\ {\bf 69} (1993) 349.}.
Proceeding loosely: we have a correspondence between reflexive polyhedra,
$\Delta$, and \cys\ $\ca{M}_\Delta$.  Combining this with the correspondence
between vector bundles $\ca{V}$ and \cys\ gives a correspondence between vector
bundles on $K3\times T^2$ and reflexive polyhedra $\ca{V}=\ca{V}_\Delta$. It
is known also that the moduli spaces of
\cys\ form a web in which
continuous transitions between different \cy\ manifolds (phases) occur
due to the shrinking to zero of certain homology two cycles
and three cycles as the parameters of the manifold are varied
\REFS{\rReid}{M.~Reid, Math.\ Ann.\ {\bf 278} (1987) 329.}
\REFSCON{\rCDLS}{P.~Candelas, A.M.~Dale, C.A.~L\"utken,
R.~Schimmrigk,\hfill\break
       \npb{298}~(1988)~493.}
\REFSCON{\rRolling}{P.~Candelas, P.S.~Green and T.~H\"ubsch,
      \npb{330} (1990) 49.}
\REFSCON{\rAGM}{P.~S.~Aspinwall, B.R.~Greene and D.R.~Morrison,\hfil\break
Int.~Math.~Res.~Notices (1993)~ 319, alg-geom/9309007.}
\refsend.
Indeed it seems likely that all \cys\ are connected by processes of this type
\REFS\rBKKII{P.~Berglund, S.~Katz and A.~Klemm, \npb{456}~(1995)~153,
hep-th/9506091.}
\REFSCON\rLS{M.~Lynker and R.~Schimmrigk, hep-th/9511058.}
\REFSCON\rACJM{A.~C.~Avram, P.~Candelas, D.~Jan\v{c}i\'{c} and M.~Mandelberg,
hep-th/9511230.}
\REFSCON\rCGGK{T.-M.~Chiang, B.~R.~Greene, M.~Gross and Y.~Kanter,
hep-th/9511204.}
\refsend.
The consistency and continuity of the string vacua associated with this
process is assured by effects involving solitons that wrap the vanishing
cycles and which become massless at the transition where the cycles vanish
\REFS{\rStro}{A.~Strominger, \npb{451}~(1995)~96, hep-th/9504090.}
\REFSCON{\rBMS}{B.~Greene, D.~R.~Morrison and A.~Strominger,\\
\npb{451}~(1995)~109, hep-th/9504145.}
\REFSCON{\rBSV}{M.~Bershadsky, V.~Sadov and C.~Vafa, hep-th/9510225.}
\REFSCON{\rKMP}{S.~Katz, D.~R.~Morrison and M.~R.~Plesser, hep-th/9601108.}
\refsend.
In virtue
of duality this web structure must exist also  on the heterotic side, though
the
physical picture is different.   Indeed, the space of heterotic vacua also
forms a web
in which different models are connected along branches parametrized by vacuum
expectation values of scalars in vector and hypermultiplets that correspond to
the
parameters of
$\ca{V}$.  In virtue of the correspondence
$\ca{V}\leftrightarrow \Delta$ there should be a dictionary that translates
between the language of reflexive polyhedra and that of heterotic
dynamics, including the non-perturbative effects uncovered by new
insight into string phenomena~
\REFS\rWit{E.~Witten, \npb{460} (1996) 541, hep-th/9511030.}
\REFSCON\rDMW{M.~Duff, R.~Minasian and E.~Witten, hep-th/9601036.}
\REFSCON\rSW{N.~Seiberg and E.~Witten, hep-th/9603003.}\refsend.
A first step towards finding such dictionary was taken in
ref.~\cite{\rAFIQ}\
where it was noticed that un-Higgsing of $SU(r)$ groups in certain
heterotic models matched with a chain of $K3$ fibrations.
Subsequently, it was argued that the appearance of perturbative
heterotic groups in these chains could be explained in the Calabi-Yau
picture as~well~\cite{\rAsp}.

Motivated by the observations of \cite{\rAFIQ},
we point out that corresponding to a Higgsing chain
of heterotic models, there is a chain of \cy\ manifolds with a
simple structure revealed by their description in terms of reflexive
polyhedra. These manifolds are $K3$ fibrations. Indeed the four-dimensional
polyhedron contains the polyhedron of the $K3$ in a simple way and it is this
nesting of polyhedra that motivates much of our analysis. One notable point is 
that the Dynkin diagram of the group and also the extended Dynkin diagram can
be seen in the edges of the $K3$ polyhedron.

The heterotic models of \cite\rAFIQ~ basically correspond to
$K3\times T^2$ compactifications in which instanton numbers
$(d_1,d_2) = (24,0), (20,4), (18,6), (16,8)$ and $(14,10)$
are embedded in $E_8 \times E_8$~
\Ref\rURA{A. Uranga, unpublished.}.
Maximally Higgsing the initial gauge group leads to models that can
be identified with type II compactifications on known $K3$
fibrations~
\REF\rMV{D.~Morrison and C.~Vafa, hep-th/9602114.}
\cite{\rKV,\rAFIQ,\rMV}.
Starting at these `irreducible models', we then study
the type II description of symmetry restoration of different
group factors. Recent results of ref.~\cite{\rMV}\ allow us to consider
other values of $(d_1,d_2)$. In some cases we find it necessary
to include the effect of $D=6$ extra tensor multiplets in the
heterotic construction \cite{\rSW}.
Reversing our strategy, we have also analyzed type II processes
that arguably correspond to non-perturbative heterotic effects.
The resulting type II pattern strongly suggests that these
processes indeed have a non-perturbative interpretation in terms
of transitions in which the number of tensor multiplets jumps
by one and an instanton shrinks to zero.

This note is organized as follows. In section 2 we analyze $K3$ and
$K3\times T^2$ heterotic compactifications, including possible
symmetry breaking
patterns. Theories of this kind have been studied in recent related work~
\REF\rVI{C.~Vafa, hep-th/9602022.}
\REF\rAFIQII{ G.~Aldazabal, A.~Font, L.~E.~Ib\'a\~nez and F.~Quevedo,
hep-th/9602097.}
\REF\rAG{P.~S.~Aspinwall and M.~Gross, hep-th/9602118.}
\cite{\rDMW,\rSW,\rMV-\rAG}.
In section 3 we consider various sequences of reflexive polyhedra
and determine their relation to heterotic models.
In section~4 we present our conclusions and list some open questions. An
appendix contains tables of Hodge numbers and figures of the polyhedra that we
discuss.
\newpage
\section{chains}{Heterotic Chains}
The starting point is a heterotic $E_8\times E_8$ compactification
on $K3$ with $SU(2)$ bundles with instanton numbers $(d_1,d_2)$ such
that $d_1+d_2=24$. When both $d_i \geq 4$, the resulting group is
the commutant of the instantons which is $E_7 \times E_7$ with massless
hypermultiplets transforming
as ${\bf 56}$s and/or singlets under each $E_7$. Using the index theorem
we find the hypermultiplet spectrum
$$
\half (d_1-4) ({\bf 56}, {\bf 1}) +
\half (d_2-4) ({\bf 1},{\bf 56}) + 62({\bf 1}, {\bf 1}) ~.$$
When $d_1=24$, $d_2=0$, the gauge group is $E_7\times E_8$
and the massless hypermultiplets include
$$ 10 ({\bf 56}, {\bf 1}) + 65({\bf 1}, {\bf 1}) ~.$$
Without loss of generality we can take $12 \leq d_1 \leq 20$, unless
$d_1=24$. We also find it convenient to define
 $$
k \define {d_1 - 12\over 2} ~.\eqlabel{kdef}$$
Since the ${\bf 56}$ of $E_7$ is a pseudoreal representation, the $d_i$
can be odd and $k$ can be half-integer.

An initial heterotic model can be deformed by vevs of hypermultiplets
thereby breaking the gauge group.
Since $d_1 \geq 12$, the number of $(\bf 56, \bf 1)$'s
is such that the first $E_7$ can be completely broken. In particular,
it can be broken through the chain
$$ E_7 \to E_6 \to SO(10) \to SU(5) \to SU(4) \to SU(3) \to
SU(2) \to SU(1) ~,
\eqlabel{higgsch}$$
where $SU(1)$ denotes the trivial group consisting of the identity only.
On the other hand, the group arising from the second $E_8$ can only
be broken to some terminal group $G^{(0)}_2$ that depends on $k$. For instance,
$G^{(0)}_2(k) = E_8, E_7, E_6, SO(8)$ for $k = 6,4,3,2$ and
$G^{(0)}_2(k) = SU(1)$ for $k=1,0$.

The hypermultiplet content at every stage of breaking can of course be
derived by group theory but it can also be found by imposing anomaly
factorization conditions. Recall that
the anomaly eight-form is given by
$$
I_8 = {1\over {16(2\pi)^4}} (\tr R^2 - v_\a \tr F_\a^2 )
(\tr R^2 - \tv_\a \tr F_\a^2 ) ~,$$
where $\a$ runs over the various gauge factors. At Kac-Moody level one,
the coefficients $v_\a$ are given by $v_\a = 2,1,{1\over 3}, {1\over 6},
{1\over 30}$ for $\a = SU(N), SO(2N), E_6, E_7, E_8$ respectively~
\Ref\rErler{J. Erler, J. Math. Phys. 35 (1993) 377, hep-th/9304104.}.
The coefficients $\tv_\a$ depend on the hypermultiplet spectrum and can
be determined from the form of the total anomaly \cite{\rErler}. For
instance,
$$
\eqalign{
\tv_{E_8} &= -{1\over 5},\crb
\tv_{SU(5)} &=  n_{10} - 2,\cr}\quad
\eqalign{
\tv_{E_7} &= {n_{56} - 4\over 6},\crb
\tv_{SU(4)}  &= n_6 - 2,\cr}\quad
\eqalign{
\tv_{E_6} &= {n_{27} -6\over 6},\crb
\tv_{SU(3)} &= {n_3 - 18\over 6},\cr}\quad
\eqalign{
\tv_{SO(10)} &= {n_{16} -4\over 2},  \crb
\tv_{SU(2)} &= {n_2 - 16\over 6},\cr} \eqlabel{vts}$$
where $n_R$ is the number of hypermultiplets in the $R$
representation. In the case of $SO(10)$, $SU(5)$, and $SU(4)$, the
number of fundamental representations is constrained to be
$n_{10} = 2 + n_{16}$, $n_5 = 10 + 3n_{10}$ and $n_4 = 8 + 4n_6$
respectively. Finally, the total number of vector multiplets, $n_V$
and hypermultiplets, $n_H$, satisfy the condition
$$ n_H - n_V = 244 ~.\eqlabel{anom}$$
It is easy to check that \eqref{anom} is satisfied in the initial
$(d_1,d_2)$ models described above.

In general, the gauge group is of the form $G = G_1\times G_2$, with
$G_1$ and $G_2$ coming from the first and second $E_8$'s. These groups are
themselves products of simple factors. Notice that before Higgsing we have
$$
{\tv_1\over v_1} =  k \quad\quad , \quad\quad
{\tv_2\over v_2} = -k ~,\eqlabel{ini}$$
where we have assumed $d_1+d_2=24$ (more generally
$\tv_2/v_2=(d_2-12)/2$).
After Higgsing, all non-Abelian factors contained in $G_2$ will
satisfy $\tv_\a/v_\a = -k$. In particular, this implies that
the terminal groups $G^{(0)}_2(k)$ mentioned above
are free of charged matter.
Similarly, all non-Abelian factors contained in
$G_1$ will satisfy $\tv_\a/v_\a = k$. For instance, if the first $E_7$
is broken to $SU(2)$, the number of doublets turns out to be
$$n_2 = 12k + 16 ~.\eqlabel{n2s}$$
The number of $SU(2)$ singlets is obtained from \eqref{anom}.
For example, if $G$ is broken to $SU(2) \times G^{(0)}_2(k)$, we find
$$n_1 = 215 + {\rm dim}\, G^{(0)}_2(k) - 24k ~.\eqlabel{n1s}$$
It is straightforward to repeat this sort of analysis for other
breaking patterns.

Up to now we have focused on six-dimensional models. Upon further
compactification on $T^2$, the $N\!=\!1$, $d\!=\!6$ hyper and vector
multiplets of $G$ give rise to $N\!=\!2$, $d\!=\!4$ hyper and vector
multiplets also of $G$, in numbers $n_H$ and $n_V$ that still must
fulfill \eqref{anom}. The tensor multiplet produces an extra $U(1)$
vector multiplet associated to the dilaton and
for generic 2-torus shape, there also appear
two extra $U(1)$ vector multiplets corresponding to
the torus that are usually denoted by $T$ and $U$. Another new feature is the
existence of a Coulomb branch parametrized by expectation values of the adjoint
scalars in the $N\!=\!2$ vector multiplets. At a generic point,
excluding the graviphoton, the
gauge group is $U(1)^{{\rm rank}\, G + 3}$ and the massless
hypermultiplets include those $n^G_{sing}$ fields originally neutral under $G$.

An $N\!=\!2$, $D\!=\!4$ heterotic model with the structure just
described is potentially equivalent to a type IIA compactification
on a \cy\ manifold that is a $K3$ fibration and has
$$\eqalign{ h_{11} & = {\rm rank}\, G + 3 \cr
h_{12} & = n^G_{sing} - 1 ~.}\eqlabel{nhs}$$
In particular, maximal Higgsing of $G$ to the matter-free
$G^{(0)}_2(k)$ leads to
 $$
\eqalign{ h_{11} & = {\rm rank}\, G^{(0)}_2(k) + 3 \cropen{5pt}
h_{12} & = 243 + {\rm dim}\, G^{(0)}_2(k)~.}\eqlabel{hter}$$
Un-Higgsing an $SU(2)$ factor in $G_1$ then changes these numbers to
 $$
\eqalign{ h_{11} & = {\rm rank}\, G^{(0)}_2(k) + 4 \cropen{5pt}
h_{12} & = 214 + {\rm dim}\, G^{(0)}_2(k) - 24k ~.}\eqlabel{h2}$$
Another interesting situation is the un-Higgsing of $SU(2)\times SU(2)$.
In this case we find
 $$\eqalign{
h_{11} & = {\rm rank}\, G^{(0)}_2(k) + 5 \cr
h_{12} & = 185 + {\rm dim}\, G^{(0)}_2(k) - 40k ~.}\eqlabel{h22}$$
Similar results can be derived for other symmetry restoration patterns.

For $k=6,4,3,2,1$, eqs.~\eqref{hter} yield Hodge numbers that
match those of known $K3$ fibrations given by hypersurfaces
of degree $12k+12$ in $\IP_4(1,1,2k,4k+4,6k+6)$
\cite{\rKV, \rAFIQ, \rMV}.
Remarkably enough, sequentially un-Higgsing $SU(r)$ factors
($r=2,\cdots , 4$) leads to Hodge numbers that also match into
those of known $K3$ fibrations \cite{\rAFIQ}. The chain of spaces thus
obtained is shown in Table~\tabref{kchain}. Each element is then
presumably equivalent to a heterotic $K3 \times T^2$ compactification
in which the gauge group can be enhanced to
$$
G = SU(r) \times G^{(0)}_2(k) \eqlabel{enhG}$$
at special points in the heterotic moduli space. Indeed, it has
been shown \cite{\rAsp}\ that such enhanced groups can also appear in
the conjectured type II dual compactification. In fact since
the toroidal $U(1)^2$ can be enhanced to $SU(3)$ there can be
overall enhancement to a group contained in $G \times SU(3)$.

\medskip

$$\vbox{
\def\skip{\hphantom{1}}
\offinterlineskip\halign{
\strut # height 15pt depth 8pt&\hfil\quad$#$\quad\hfil\vrule
&\quad$#$\quad\hfil\vrule&\quad$#$\quad\hfil\vrule  \cr
\noalign{\hrule}
\vrule&\hfil r&\hfil \hbox{\cy\ Manifold} & \hfil K3\cr
\noalign{\hrule\vskip3pt\hrule}
\vrule & 4
        & \cp5^{(1,1,2k,2k{+}4,2k{+}6,2k{+}8)}[4k{+}8, 4k{+}12]
         & \cp4^{(1,k,k{+}2,k{+}3,k{+}4)}[2k{+}4, 2k{+}6]\cr
\vrule & 3
        & \cp4^{(1,1,2k,2k{+}4,2k{+}6)}[\skip 6k{+}12]
         & \cp3^{(1,k,\skip k{+}2,\skip k{+}3)}[3k{+}6]\cr
\vrule & 2
        & \cp4^{(1,1,2k,2k{+}4,4k{+}6)}[\skip 8k{+}12]
         & \cp3^{(1,k,\skip k{+}2,2k{+}3)}[4k{+}6]\cr
\vrule & 1
        & \cp4^{(1,1,2k,4k{+}4,6k{+}6)}[12k{+}12]
         & \cp3^{(1,k,2k{+}2,3k{+}3)}[6k{+}6]\cr
\noalign{\hrule}
}}
$$
\nobreak\tablecaption{kchain}{The $k$'th chain of hypersurfaces with
enhanced group $SU(r)\times G^{(0)}_2(k)$, $k=1, \cdots , 6$.}
\bigskip

The case $k=5$ is special in that the chain of hypersurfaces do not in fact
correspond to heterotic models of the type we have described since this
would require $d_2=2 < 4$. Moreover, these hypersurfaces do not appear
in the lists~\cite{\rKleSch}
of \cy\ spaces in weighted $\IP_4$ owing to the fact that
the weights do not allow for transverse polynomials. This second objection is,
however, easily dealt with.
It is possible to give meaning to the manifolds of this
chain in virtue of the correspondence between \cys\ and reflexive polyhedra
that is provided by the construction of Batyrev \cite{\rBat}\ in terms of
reflexive polyhedra and it is therefore tempting to include this chain here.
As for the first objection, we shall see
in the next section, that the structure of the $K3$
polyhedra in the $k=5$ chain is identical to that in the $k=6$ chain.
This strongly suggests that the terminal group is $G^{(0)}_2(5)=E_8$.
This proposal works if we modify the $N=1,~D=6$ construction so as to include
effects seen in the compactification of $M$-theory
\REFS\rHW{P.~Horava and E.~Witten, \npb{460} (1996) 506, hep-th/9510209.}
\REFS\rWitII{E.~Witten, hep-th/9512219.}
\cite{\rHW, \rWitII, \rDMW, \rSW}
and $F$-theory \cite{\rVI,\rMV}.
More precisely we consider vacua with $n_T$ tensor multiplets
so that the condition $d_1+d_2=24$ is replaced by
$$
d_1 + d_2 + n_T - 1 =24~.\eqlabel{newinst}$$
We immediately see that $n_T=3$ permits $d_1=22,~d_2=0$.
In this way we can have a $k=5$ chain. Moreover, the initial gauge group
$E_7 \times E_8$ can be completely broken to a matter-free $E_8$.

The expressions for $h_{11}$ and $h_{12}$ must be modified
since owing to the tensor multiplets, eq.~\eqref{anom} becomes \cite{\rSW}
$$
n_H-n_V = 273 - 29n_T\eqlabel{newanom}$$
and we must also take into account that upon further compactification on $T^2$
the $n_T$ tensor multiplets give rise to $n_T$ $U(1)$ vector multiplets.
The effect is that eqs.~\eqref{hter} are now replaced by
$$
\eqalign{ h_{11} & = {\rm rank}\, G^{(0)}_2(k) + n_T + 2 \cr
h_{12} & = 272 + {\rm dim}\, G^{(0)}_2(k) -29n_T ~.}\eqlabel{htern}$$
Eqs.~\eqref{h2} and \eqref{h22} change accordingly. With
$n_T=3$ and $G^{(0)}_2(5)=E_8$ we recover the observed Hodge numbers
of the $k=5$ spaces.

For $k=0$ we have a heterotic construction but this value does not admit of a
simple interpretation in terms of the manifolds of Table~\tabref{kchain}.
However, this interesting case can be
analyzed following the approach of ref.~\cite{\rMV}. Indeed, it has been
pointed out that the terminal hypersurface
$\cp4^{(1,1,2k,4k{+}4,6k{+}6)}[12k{+}12]$ can also be viewed as an
elliptic fibration over the Hirzebruch surface $\IF_{2k}$ \cite{\rMV}.
As we will explain shortly, in this setting it is natural to consider
$k=0$ as well as $k=$half-integer. We will now briefly discuss
the latter situation.

When $k=1/2$, the second $E_7$ can be broken completely so that
$G^{(0)}_2(1/2) = SU(1)$. When $k=3/2$, sequential breaking ends in
$SU(3)$ \cite{\rDMW}.
Notice that $\tv/v = -3/2$ for a matter-free $SU(3)$.
When $k=5/2$ the breaking can proceed to $G^{(0)}_2(5/2)=F_4$~
\Ref\rVK{V.~Kaplunovsky, private communication.}.
For a
matter-free $F_4$, $\tv/v = -5/2$ \cite{\rErler}, as expected. When
$k=7/2$, there is no known breaking, the terminal group is just $E_7$
with a half ${\bf 56}$ hypermultiplet. Finally, the values
$k=9/2, 11/2$ require $n_T=4,2$ tensor multiplets and since $d_2=0$,
the terminal group is~$E_8$. This is summarized by Table~\tabref{terminal}.

In general, if $G_2$ stays completely broken at a matter free
$G_2^{(0)}(k)$ and considering different $G_1 \to H$ breaking
patterns leads to models with Hodge numbers of the form
 $$\eqalign{
h_{11} &= {\rm rank}\, G_2^{(0)}(k) + {\rm rank}\,H + n_T(k) + 2\cropen{5pt}
h_{21} &= 272 + {\rm dim}\,G_2^{(0)}(k) + {\rm dim}\,H - 29n_T(k)
- a_H - b_H k~.\cr} \eqlabel{hodgeformulas}$$
The coefficients $a_H$ and $b_H$ encode the number of $H$-charged
fields that disappear in the Coulomb phase, their values are recorded in
Table~\tabref{bottoms} for a number of groups that we shall meet.
Although $SO(12)$ cannot be obtained from $E_7$ by the usual
Higgs mechanism, it arises naturally from breaking of the
original $E_8$ by background $SU(2)\times SU(2)$ fields with
total instanton number $d_1=12+2k$. For $SU(6)$ we have considered
the two simplest possibilities with $\tv/v = k$. For $SU(6)$,
$n_6=16+4k$, $n_{15}=2k+2$ and $n_{20}=0$, whereas for $SU(6)_b$,
$n_6=18+6k$, $n_{15}=0$ and $n_{20}=k+1$ (the {\bf 20} is a pseudoreal
representation). Each of these cases can be obtained by Higgsing an
$SO(12)$ with appropriate numbers of ${\bf 32}$s
and ${\bf 32^\prime}$s that
in turn depend on how $d_1$ is distributed between the two $SU(2)$'s~\
\Ref\rANGEL{A.~Uranga, private communication.}.
\bigskip
$$
\def\box#1{\hbox to 25.9pt{\hfil$#1$\hfil}\vrule}
\vbox{\offinterlineskip\halign{
\strut # height 15pt depth 8pt&\hskip5pt\hfil$#$\hfil\hskip5pt
\vrule\hskip1pt\vrule
&\box{#}&\box{#}&\box{#}&\box{#}&\box{#}&\box{#}
&\box{#}&\box{#}&\box{#}&\box{#}&\box{#}&\box{#}&\box{#}
\cr
\noalign{\hrule}
\vrule&k&0&{1\over 2}&1&{3\over 2}&2&{5\over 2}&3&{7\over 2}&4
&{9\over 2}&5&{11\over 2}&6\cr
\noalign{\hrule}
\vrule&G_2^{(0)}&SU_1&SU_1&SU_1&SU_3&SO_8&F_4&E_6
&E_7^{-}&E_7&E_8&E_8&E_8&E_8\cr
\noalign{\hrule}
\vrule&\hbox{rk}[G_2^{(0)}]&0&0&0&2&4&4&6&7&7&8&8&8&8\cr
\noalign{\hrule}
\vrule&\!\hbox{dim}[G_2^{(0)}]\!&0&0&0&8&28&52&78&105&133&248&248&248&248\cr
\noalign{\hrule}
\vrule&n_T&1&1&1&1&1&1&1&1&1&4&3&2&1\cr
\noalign{\hrule}
}}
$$
\nobreak\tablecaption{terminal}{\baselineskip=13pt Terminal groups and the
numbers of tensor multiplets for the different values of $k$.
The entry $E_7^-$ corresponding to $k={7\over 2}$ denotes $E_7$ with a
half-multiplet of {\bf 56}. The entries in this column give the values
that are appropriate to the use of the relations \eqref{hodgeformulas}
for this case.}
\newpage
\section{polyhedra}{Sequences of Reflexive Polyhedra}
\vskip-20pt
\subsection{General structure}
We now set out to study the \cy\ spaces conjectured to give the type IIA
dual description of the heterotic compactifications explained in the
previous section. Our strategy is to analyze the manifolds following
Batyrev's toric approach. (For a concise summary of Batyrev's
construction in a form accesible to physicists see, for example,~\cite{\rCDK}.)
To a \cy\ manifold (of any dimension) defined as a hypersurface in a
weighted projective space one can associate its Newton polyhedron,
which we denote by $\D$. The Newton polyhedron is often (perhaps always)
reflexive and when it is we may define the
dual or polar polyhedron which we denote by $\nabla$.
By means of a computer program we have computed the dual polyhedra for
the last three manifolds given in Table~\tabref{kchain}.
The polyhedra for each $k$ have similar properties that we will
illustrate by considering $k=1$ as an example. For this case, the
points of the dual polyhedra are displayed in Table~\tabref{chainone}.
We shall modify these polyhedra shortly so for the present the
$\nabla$'s are distinguished by tildes.
\bigskip
 $$\vbox{\offinterlineskip\halign{\strut # height 12pt depth 5pt
&\hfil\quad$#$\quad\hfil\vrule
&\hfil\quad$#$\quad\hfil\vrule
&\hfil\quad$#$\quad\hfil\vrule \cr
\noalign{\hrule}
\vrule&{}^4\widetilde\nabla^{SU(1)}
       &{}^4\widetilde\nabla^{SU(2)}
        &{}^4\widetilde\nabla^{SU(3)}\cr
\noalign{\hrule\vskip3pt\hrule}
\omit{\vrule height2pt}&&&\cr
\vrule&( -1,\- 0,\- 2,\- 3)   &( -1,\- 0,\- 2,\- 3)   &(  -1,\- 0,\- 2,\- 3)\cr
\vrule&\llap{*}\,(\- 0, -1,\- 2,\- 3)   &\llap{*}\,(\- 0, -1,\- 1,\- 2)
       &\llap{*}\,(\- 0, -1,\- 1,\- 1)\cr
\vrule&(\- 0,\- 0, -1,\-  0)  &(\- 0,\- 0,  -1,\- 0)  &(\- 0,\- 0,  -1,\- 0)\cr
\vrule&(\- 0,\- 0,\- 0,  -1)  &(\- 0,\- 0,\- 0,  -1)  &(\- 0,\- 0,\- 0,  -1)\cr
\vrule&(\- 0,\- 0,\- 0,\- 0)  &(\- 0,\- 0,\- 0,\- 0)  &(\- 0,\- 0,\- 0,\- 0)\cr
\vrule&(\- 0,\- 0,\- 0,\- 1)  &(\- 0,\- 0,\- 0,\- 1)  &(\- 0,\- 0,\- 0,\- 1)\cr
\vrule&(\- 0,\- 0,\- 1,\- 1)  &(\- 0,\- 0,\- 1,\- 1)  &(\- 0,\- 0,\- 1,\- 1)\cr
\vrule&(\- 0,\- 0,\- 1,\- 2)  &(\- 0,\- 0,\- 1,\- 2)  &(\- 0,\- 0,\- 1,\- 2)\cr
\vrule&(\- 0,\- 0,\- 2,\- 3)  &(\- 0,\- 0,\- 2,\- 3)  &(\- 0,\- 0,\- 2,\- 3)\cr
\vrule&(\- 0,\- 1,\- 2,\- 3)  &(\- 0,\- 1,\- 2,\- 3)  &(\- 0,\- 1,\- 2,\- 3)\cr
\vrule&(\- 1,\- 2,\- 2,\- 3)  &(\- 1,\- 2,\- 2,\- 3)  &(\- 1,\- 2,\- 2,\- 3)\cr
\omit{\vrule height2pt}&&&\cr
\noalign{\hrule}
}}
$$
\nobreak\tablecaption{chainone}{The dual polyhedra for $k=1$ and
$H=SU(1),\,SU(2)$ and $SU(3)$.}
\bigskip

\noindent
The following observations summarize the structure of the polyhedra:

\item{1.~~}For each polyhedron all the points except two (the first and the
last) lie in the hyperplane $x_1=0$. In each case the points that lie in the
hyperplane $x_1=0$ themselves form a reflexive polyhedron which is
the dual of the polyhedron for the respective $K3$ of the fibration.

\item{2.~~}Omitting the first two points and the last two points of each
polyhedron leaves us with a two-dimensional
reflexive polyhedron, $\nab{2}$, which is a triangle.
This $\nab{2}$ is the dual polyhedron of the torus $\cp2^{(1,2,3)}[6]$.
We see in this way that the $K3$'s are elliptically fibred as elucidated in
ref.~\cite{\rMV}

\item{3.~~}The three polyhedra of Table~\tabref{chainone} differ only in the
second point of each (which is distinguished by an asterisk).
Let $\d$ denote the (non-reflexive) polyhedron consisting of the common
points and denote by $pt'_r,~r=1,2,3,$ the three special points
$pt'_1 = (0,-1,2,3)$, $pt'_2 = (0,-1,1,2)$ and $pt'_3 = (0,-1,1,1)$ (the
utility of the primes will become apparent as we proceed). Then we have
 $$
{}^4\widetilde\nabla^{SU(r)} = \d\cup pt'_r~.$$
We observe that we can add points to the polyhedra as follows without changing
the
Hodge numbers of the associated manifolds.
 $$\eqalign{
{}^4\nabla^{SU(1)} &= {}^4\widetilde\nabla^{SU(1)} \cr
{}^4\nabla^{SU(2)} &= {}^4\widetilde\nabla^{SU(2)}\cup pt'_1 \cr
{}^4\nabla^{SU(3)} &= {}^4\widetilde\nabla^{SU(3)}\cup pt'_1\cup pt'_2 \cr}$$
The fact that the Hodge numbers of ${}^4\nabla^{SU(r)}$ and
${}^4\widetilde\nabla^{SU(r)}$ are the same may mean that the polyhedra
correspond to the same manifold. However, whether this is or not the case,
we will take the sequence of polyhedra on the left of these
relations as defining the chain, abandoning if need be the sequence
of spaces of Table~\tabref{kchain} that was our original motivation.

\item{4.~~}The $pt'_r$ lie in the hyperplane $x_1=0$ and the observation
above that the points of the polyhedra that lie in the
hyperplane $x_1=0$ form a reflexive polyhedron continues to
hold for the augmented polyhedra ${}^4\nabla^{SU(r)}$.
We are therefore dealing with a succession of $K3$
manifolds and the relations above hold equally well if each ${}^4\nabla$
is replaced by a ${}^3\nabla$ referring to the $K3$'s.

\item{5.~~}Fix attention now on the polyhedra, ${}^3\nabla$, of
the $K3$'s and the two dimensional polyhedron, ${}^2\nabla$, of the
torus. The ${}^2\nabla$ is the same for each of the three spaces. It
consists of the 7 points that have $x_1 = x_2 = 0$. ${}^2\nabla$ divides
${}^3\nabla$ into a `top', ${}^3\nabla_{\hbox{top}}$, consisting of
points for which $x_1=0$ and $x_2\ge 0$
and a `bottom', ${}^3\nabla_{\hbox{bot}}$, consisting of points for which
$x_1=0$ and $x_2\le 0$. Figure~\figref{nabla} illustrates this for the
case $k=3$ and $r=3$. As we move up the chain
${}^3\nabla_{\hbox{bot}}$ changes, it is succesively
 $$
{}^2\nabla\cup pt'_1 \to {}^2\nabla\cup pt'_1\cup pt'_2 \to
{}^2\nabla\cup pt'_1\cup pt'_2\cup pt'_3 \to \cdots~. \eqlabel{sequ}$$
The polyhedron ${}^3\nabla_{\hbox{top}}$
however is unchanged as we move up the chain.

\midinsert
\def\nablathreethree{
\vbox{\vskip0pt\hbox{\epsfxsize=6truein\epsfbox{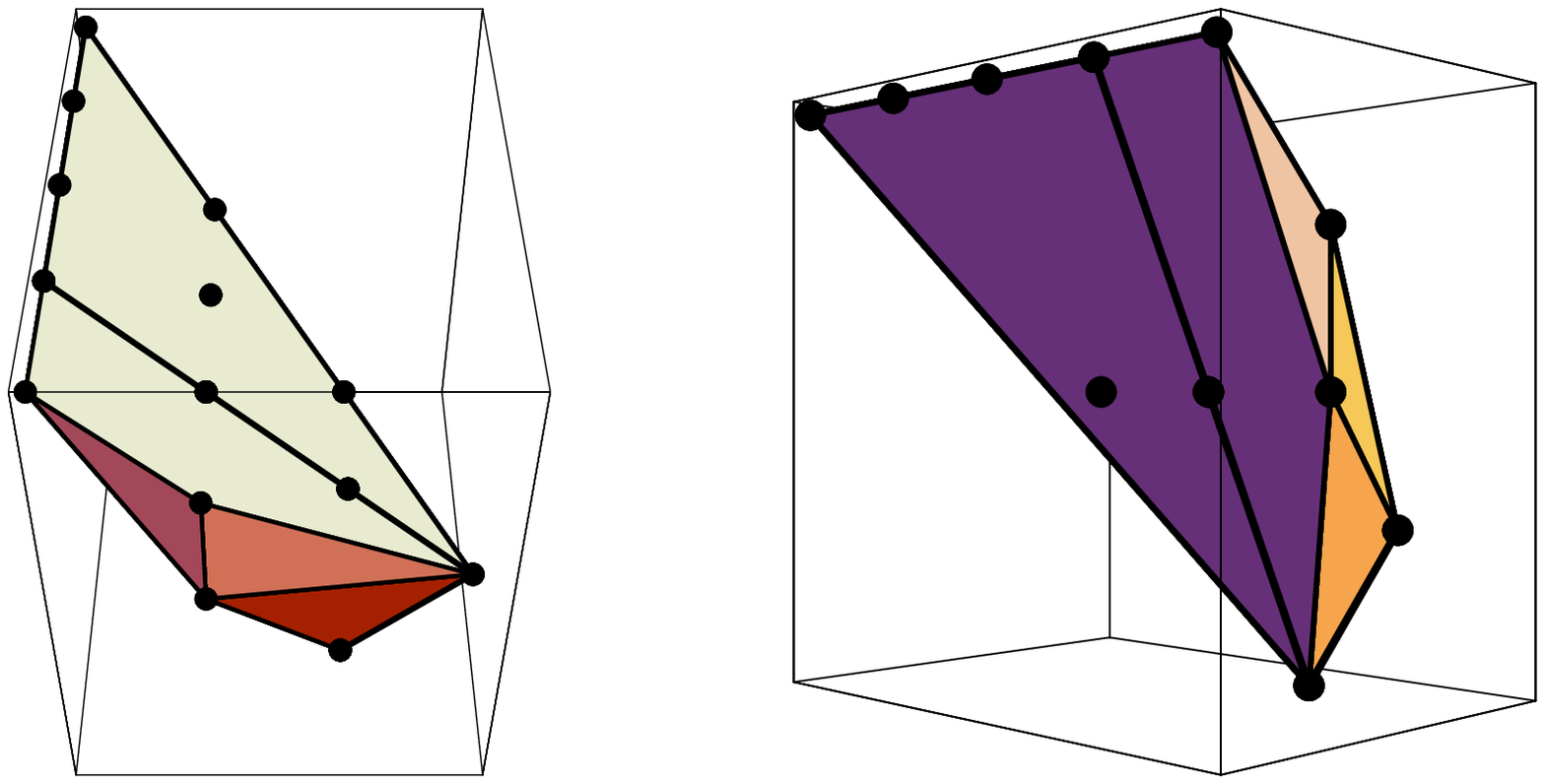}}}}
\figbox{\nablathreethree\vskip0pt}{\figlabel{nabla}}
{Two views of the polyhedron $^3\nabla^{k=3,\, SU(3)}$. The polyhedron is
divided into a top and a bottom by the triangle $\nab{2}$.}
\place{1.2}{2.7}{$\nab{2}$}
\endinsert

\noindent
We can now elaborate on our statement that for each chain a similar
pattern obtains. Apart from two points, which for each member of the $k$'th
chain are $(-1, 0, 2, 3)$ and $(1, 2k, 2, 3)$, the points of the polyhedron
lie in the plane $x_1 = 0$ forming the polyhedron, $\nab{3}$, of the $K3$.
For each member of a chain, the polyhedron of the $K3$ is again divided
into a top and a bottom by the  polyhedron
$\nab{2}$ of the torus and we may write
 $$\nab{3}^{k,H} = \nabla_{\hbox{bot}}^H\cup\nabla_{\hbox{top}}^k ~,
\eqlabel{nk3}$$
where $\nabla_{\hbox{top}}^k$ depends only on $k$ while
$\nabla_{\hbox{bot}}^H$ depends only on the group $H$ that is
perturbatively restored in the heterotic side.
In particular then, the dual polyhedron ${}^4\nabla^{k,SU(1)}$ of the
lowest space $\cp4^{(1,1,2k,4k{+}4,6k{+}6)}[12k{+}12]$ can be written as
$${}^4\nabla^{k,SU(1)} = \nab{3}^{k,SU(1)} \cup
\{ (-1,0,2,3),\,(1,2k,2,3)\} ~. \eqlabel{nose}$$

We can describe the tops and bottoms of $\nab{3}^{k,H}$ quite simply.
If we denote by $\ca{T}^k$ the tetrahedron with base $\nab{2}$ and top vertex
$(0,k,2,3)$, the top polyhedron may be specified as follows:
 $$\eqalign{
\nabla_{\hbox{top}}^k &= \ca{T}^k~,~~\hbox{for}~k=1,2,3\cr
\nabla_{\hbox{top}}^4 &= \ca{T}^4\cup \{ (0,1,0,0),\,(0,2,0,1),\,(0,3,1,2)
\}\cr
\nabla_{\hbox{top}}^5 &= \nabla_{\hbox{top}}^6 = \ca{T}^6~.\cr}
$$

\midinsert
\def\torus{
\vbox{\vskip10pt\hbox{\epsfxsize=2.0truein\epsfbox{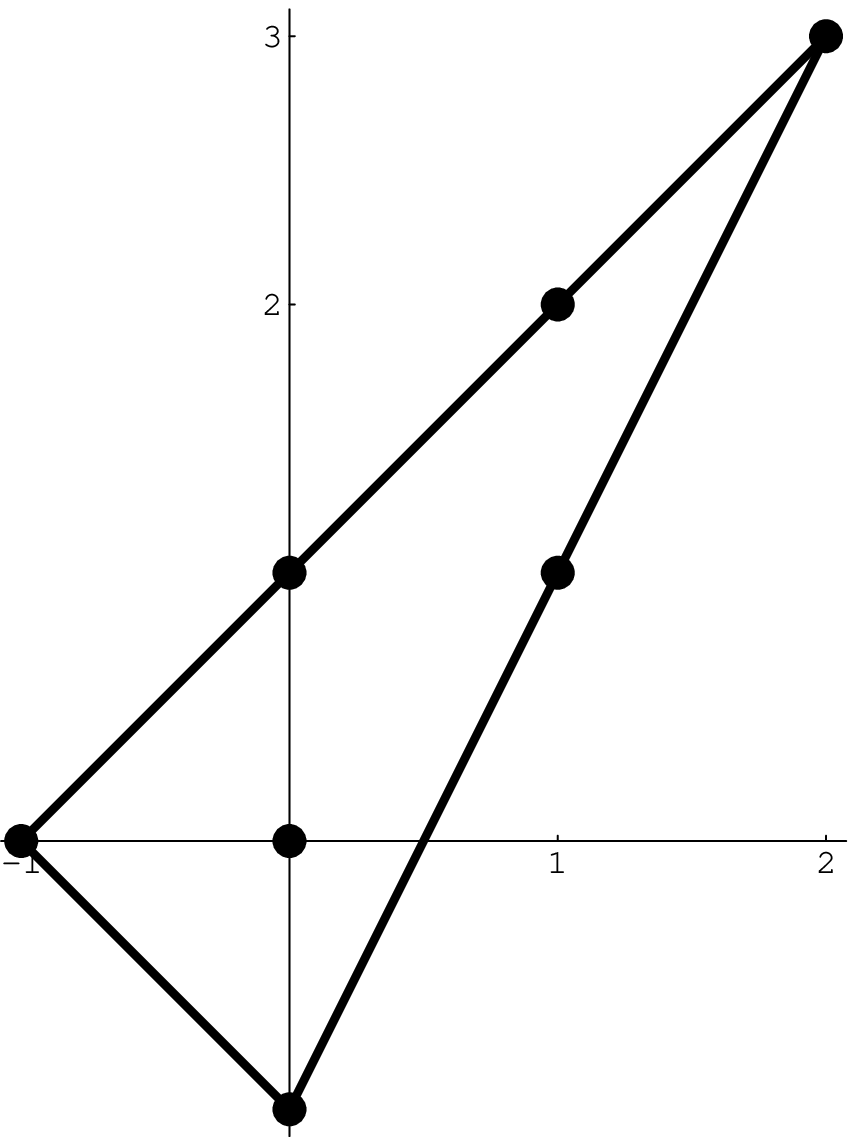}}}}
\figbox{\torus\vskip0pt}{\figlabel{torus}}
{The polyhedron, $^2\nabla$, of $\cp2^{(1,2,3)}[6]$. The
$pt'_r$ are the points directly below the indicated points of
the plot.}
\place{4.05}{3.7}{$pt_1$}
\place{3.4}{3.1}{$pt_2$}
\place{3.7}{2.3}{$pt_3$}
\place{2.7}{2.45}{$pt_4$}
\place{2.75}{1.5}{$pt_5$}
\place{2.05}{1.8}{$pt_6$}
\place{3.1}{1.05}{$pt_7$}
\endinsert
\pageinsert
$$
\def\skip{\hskip4pt}
\vbox{\offinterlineskip\halign{
\strut # height 12pt depth 5pt&\quad $#$ \skip \hfil\vrule
&\hskip10pt  $#$ \skip \hfil\vrule
&\quad$#$\quad\hfil\vrule&\quad$#$\quad\hfil\vrule&\quad$#$\quad\hfil\vrule\cr
\noalign{\hrule}
\vrule&\hfil H&\hfil\hbox{Bottom}&\hfil a_H&\hfil b_H&\hfil\tilde{n}_T\cr
\noalign{\hrule\vskip3pt\hrule}
\vrule&SU(1) &\{ pt_1' \}&0&0&n_T\cr
\vrule&SU(2) &\{ pt_1', pt_2' \}&32&24&n_T\cr
\vrule&SU(3) &\{ pt_1', pt_2', pt_3' \}&54&36&n_T\cr
\vrule&G_2   &\{ pt_1'', pt_2', pt_3' \}&60&36&n_T\cr
\vrule&SO(5) &\{ pt_1', pt_2', pt_4' \}&68&40&n_T\cr
\vrule&SU(4) &\{ pt_1', pt_2', pt_3', pt_4' \}&76&44&n_T\cr
\vrule&SO(7) &\{ pt_1'', pt_2', pt_3', pt_4' \}&82&44&n_T\cr
\vrule&Sp_3 &\{ pt_1', pt_2', pt_4', pt_6' \}&108&48&n_T\cr
\vrule&SU(5) &\{ pt_1', pt_2', pt_3', pt_4', pt_5' \}&100&50&n_T\cr
\vrule&SO(9) &\{ pt_1'', pt_2'', pt_3', pt_4' \}&104&48&n_T\cr
\vrule&F_4   &\{ pt_1''', pt_2'', pt_3', pt_4' \}&120&48&n_T\cr
\vrule&SU(6)&\{ pt_1', pt_2', pt_3', pt_4', pt_5', pt_6' \}&126&54&n_T\cr
\vrule&SU(6)_b&\{ pt_1', pt_2', pt_3', pt_4', pt_5', pt_7' \}&128&56&n_T\cr
\vrule&SO(10)&\{ pt_1'', pt_2'', pt_3', pt_4', pt_5' \}&124&52&n_T\cr
\vrule&SO(11)&\{ pt_1'', pt_2'', pt_3', pt_4'', pt_5' \}&134&52&n_T\cr
\vrule&SO(12) &\{ pt_1'', pt_2'', pt_3', pt_4'', pt_5', pt_6'  \}&160&56&n_T\cr
\vrule&E_6   &\{ pt_1''', pt_2'', pt_3'', pt_4', pt_5' \}&162&54&n_T\cr
\vrule&E_7   & \{ pt_1'''', pt_2''', pt_3'', pt_4'', pt_5' \}&224&56&n_T\cr
\noalign{\hrule\vskip3pt\hrule}
\vrule&SU(6)_c&\{ pt_1', pt_2', pt_3', pt_4', pt_5', pt_6', pt_7'\}&72&0&n_T+2k+2
\cr
\vrule&SO(13) &\{ pt_1'', pt_2'', pt_3', pt_4'', pt_5', pt_6'' \}&60&0&n_T+2k+4\cr
\vrule&E_{6B} &\{ pt_1''', pt_2'', pt_3'', pt_4', pt_5', pt_7' \}&0&0&n_T+2k+6\cr
\vrule&E_{7B}&\{ pt_1'''', pt_2''', pt_3'', pt_4'', pt_5', pt_6' \}&0&0&
n_T+2k+8\cr
\vrule&E_8 &\{ pt_1^{(6)}, pt_2^{(4)}, pt_3''', pt_4'', pt_5'\}&0&0&n_T+2k+12\cr
\noalign{\hrule}
}}
$$
\tablecaption{bottoms}{\baselineskip=13pt The bottoms containing $pt'_1$ formed
by adding the points $pt_r^{(j)}$ to $\nab2$.
In each case the points of $\nab2$ are understood and the points that are
written are the lowest members of columns. Thus $pt'''_2$ for example implies the
presence of $pt''_2$ and $pt'_2$. The bottoms that appear in the lower block
correspond to nonperturbatively realised groups. The coefficients
$a_H$ and $b_H$, on the right, are the quantities that appear in the expressions
\eqref{hodgeformulas} for the Hodge numbers. The quantity $\tilde{n}_T$ denotes the
number of tensor multiplets while $n_T$ refers to the quantities given by the third
row of Table~\tabref{terminal}.}
\vfil
\endinsert
\topinsert
$$
\def\skip{\hskip4pt}
\vbox{\offinterlineskip\halign{
\strut # height 12pt depth 5pt&\quad $#$ \skip \hfil\vrule
&\hskip10pt  $#$ \skip \hfil\vrule
&\quad$#$\quad\hfil\vrule&\quad$#$\quad\hfil\vrule&\quad$#$\quad\hfil\vrule\cr
\noalign{\hrule}
\vrule&\hfil H&\hfil\hbox{Bottom}&\hfil a_H&\hfil b_H& \hfil\tilde{n_T}\cr
\noalign{\hrule\vskip3pt\hrule}
\vrule&SU(2)_b &\{ pt_2' \}&32&24&n_T\cr
\vrule&SU(2)_c &\{ pt_4' \}&62&36&n_T\cr
\vrule&SU(2)_d &\{ pt_6' \}&92&36&n_T\cr
\vrule&SU(2)\times SU(2) &\{ pt_2', pt_4' \}&64&40&n_T\cr
\vrule&(SU(2)\times SU(2))_b &\{ pt_4', pt_6' \}&94&44&n_T\cr
\vrule&SU(3)\times SU(2) &\{ pt_2', pt_4', pt_5' \}&86&48&n_T\cr
\vrule&(SU(3)\times SU(2))_b &\{ pt_3', pt_5' \}&86&48&n_T\cr
\vrule&(SU(3)\times SU(2))_c &\{ pt_5' \}&86&48&n_T\cr
\vrule&SO(5)\times SU(2) &\{ pt_2', pt_4', pt_6' \}&100&48&n_T\cr
\vrule&G_2\times SU(2) &\{ pt_2', pt_4'', pt_5' \}&92&48&n_T\cr
\vrule&SU(4)\times SU(2) &\{ pt_2', pt_4', pt_5', pt_6' \}&108&52&n_T\cr
\vrule&SO(7)\times SU(2) &\{ pt_2', pt_4'', pt_5', pt_6' \}&114&52&n_T\cr
\vrule&SO(9)\times SU(2) &\{ pt_2', pt_4'', pt_5', pt_6'' \}&136&56&n_T\cr
\vrule&&&&&\cr
\vrule&SU(3)_b &\{ pt_3' \}&54&36&n_T\cr
\vrule&SU(3)_c &\{ pt_7' \}&102&48&n_T\cr
\vrule&SU(3)\times SU(3) &\{ pt_3', pt_5', pt_7' \}&108&54&n_T\cr
\noalign{\hrule\vskip3pt\hrule}
\vrule&G_2\times SU(3)   &\{ pt_3', pt_5', pt_7'' \}&114&0&n_T+2k\cr
\vrule&F_4\times SU(2) &\{ pt_2', pt_4'', pt_5', pt_6''' \}&152&0&n_T+2k\cr
\noalign{\hrule}
}}
$$
\tablecaption{bottomstwo}{\baselineskip=13pt 
The bottoms that do not contain $pt'_1$. The bottoms that are given in the lower
block again correspond to groups that are realised nonperturbatively.}
\endinsert
We can also easily describe $\nabla_{\hbox{bot}}^H$
for $H=SU(r)$, $r=1,2,3$. Consider first the polyhedron
$\nab{2}$ shown in Figure~\figref{torus} and let $pt'_r$ be,
as previously, the points of the lattice that are directly
below the corresponding points of $\nab{2}$.
We find that
 $$\nabla_{\hbox{bot}}^{SU(r)} =\nab{2}\cup \bigcup_{j=1}^r pt'_j ~.
\eqlabel{sur}$$
Moreover, we have verified that including the point $pt'_4=(0,-1,0,1)$
leads to a reflexive polyhedron that gives the expected Hodge numbers
for an enhanced $SU(4)$. Further adding $pt'_5=(0,-1,0,0)$ corresponds
to an enhanced $SU(5)$ and $pt'_6= (0,-1,-1,0)$ to an enhanced $SU(6)$. Hence,
eq.~\eqref{sur} is actually valid for
$r=1,\cdots ,6$. Note that we are abandoning the $SU(4)$ manifolds of
Table~\tabref{kchain} in favour of the definition that we are giving here. It
would be of interest to see if these manifolds are in fact the same.

We denote by $pt''_r$ the point that is two levels vertically below $pt_r$
and more generally by $pt_r^{(j)}$ the point that is $j$ levels below.
We consider now the possibility of adding combinations of the points
$pt_i^{(j)}$ in all possible ways such that the bottom corresponds to a
reflexive polyhedron that contains $pt'_1$.
This is straightforward in virtue of the fact that
if the polyhedron is to be reflexive the point $pt'_5$ cannot be
interior to the polyhedron. Since the polyhedra must be convex
we can at most drop down 6 levels below $pt_1$, 4 below $pt_2$, 3 below
$pt_3$, 2 below $pt_4$, 1 below $pt_5$ and 2 below each of $pt_6$ and
$pt_7$.  Very few of the possible combinations lead to convex
polyhedra. We have enumerated all the combinations of points that do.
Table~\tabref{bottoms} shows the allowed bottoms leading to
reflexive polyhedra that correspond to
enhanced gauge groups $H\times G_2^{(0)}(k)$. The resulting Hodge numbers,
recorded in
Tables~A\tabref{Hodge} and A\tabref{moreHodge} in the Appendix, can
be seen to agree with \eqref{hodgeformulas}.

All the groups obtained by the sequential Higgsing \eqref{higgsch}
do appear and, as a matter of consistency, we see the inclusions
$$
E_7\supset E_6\supset SO(10)\supset SU(5)\supset SU(4)\supset SU(3)
\supset SU(2)\supset SU(1) ~,$$
as inclusions of the respective polyhedra. We also observe that
$SO(10)\supset SO(8)\supset SU(4)\supset SU(2)^2$ as expected
from Higgsing. Likewise, there are bottoms corresponding to
$SO(12)\supset SU(6)\supset SU(2)^3$ that contain $pt'_6$ which is
not a point of the $E_7$ polyhedron. Finally, only the bottom for
$SU(6)_b$ contains $pt'_7$. Clearly $SU(6)_b\supset SU(5)$.
The fact that $SO(12) \not\supset SU(6)_b$ suggests that
the matter content of this $SO(12)$ does not include
${\bf 32}$s that could give rise to the ${\bf 20}$s present
in $SU(6)_b$.

The rest of the bottoms giving reflexive polyhedra are presented in the lower 
block of Table~\tabref{bottoms}. Before discussing their interpretation, which
involves non-perturbative effects, we describe the extension of the previous
analysis to the remaining values of $k$.
\subsection{$k=0$ and $k=$~half integer}
We wish to consider also the cases for which $k=0$ and $k=$~half-integer.
It is simplest to discuss first the cases
$k={3\over 2},{5\over 2},\ldots ,{11\over 2}$.
For these cases the manifolds in the left hand column of Table~\tabref{kchain}
still make sense though the $K3$-fibration is less easy to see. On
constructing the duals of the Newton polyhedra we find that precisely the
same structure emerges as for the case of $k$ integral. In particular
all points except two lie in a plane and these points form the
polyhedron, $\nab{3}$, of a $K3$. This shows that the
manifolds are indeed $K3$-fibrations with generic fiber corresponding to the
polyhedron $\nab{3}$. Again each $\nab{3}$ contains a $\nab{2}$ which
divides the
$\nab{3}$'s into tops and bottoms with the bottoms independent~of~$k$.

For $k=0$ the manifolds of Table~\tabref{kchain} make no sense since the
weight of
the third coordinate would be zero. While for $k=\half$ the lowest member of
the
chain would be $\cp{4}^{(1,1,1,6,9)}[18]$ which is not a $K3$-fibration.
The cases $k=0$ and $k=\half$ are however covered by a construction of
Morrison and Vafa~\cite{\rMV} which
realizes the lowest member of each chain as an elliptic
fibration over the Hirzebruch surface $\IF_{2k}$. Each manifold is
described as a
space of seven complex variables $s,t,u,v,x,y,z$ subject to
three scaling symmetries with parameters $\l,\m,\n$.
The variables scale with weights shown in Table~\tabref{scaling}.
\bigskip
\vbox{\offinterlineskip
$$\vbox{\offinterlineskip\halign{
&\strut\vrule height 12pt depth 6pt #&\hfil\quad$#$\quad\vrule
&\hfil \quad$#$\quad&\hfil\qquad$#$\quad&\hfil\qquad$#$\quad
&\hfil\qquad$#$\quad&\hfil\qquad$#$\quad&\hfil\qquad$#$\quad
&\hfil\qquad$#$\quad\vrule&\quad$#$\quad\hfil\vrule\cr
\noalign{\hrule}
&&s&t&u&v&x&y&z&\hfil\hbox{degrees}\cr
\noalign{\hrule\vskip3pt\hrule}
&\l&1&1&\hidewidth{2k}&0&\hidewidth{4k{+}4}&\hidewidth{6k{+}6}&0&12k+12\cr
&\m&0&0&1&1&4&6&0&12\cr
&\n&0&0&0&0&2&3&1&6\cr
\noalign{\hrule}
}}
$$
\tablecaption{scaling}{The scaling weights of the elliptic
fibration over $\IF_{2k}$.}
}
\bigskip

The Newton Polyhedron of the \cy\ manifold defined by the data in
Table~\tabref{scaling} is constructed by finding all possible
monomials $s^{m_1}t^{m_2}u^{m_3}v^{m_4}x^{m_5}y^{m_6}z^{m_7}$ whose degrees
under the three scalings are given by the right hand column of the Table.
Owing to the three constraints on the seven exponents, $m_1,\ldots, m_7$,
the allowed monomials lie in a four-dimensional lattice within $\IR^7$. The
convex hull of the monomials is the Newton polyhedron. The polyhedron contains
the point $(1,1,1,1,1,1,1)$, corresponding to the monomial $stuvxyz$, as its
only interior point. By translating the origin to this point, constructing
the dual
polyhedron and making a $GL(4,\IZ)$ transformation we obtain, in all cases,
polyhedra that are  exactly of the form given
in \eqref{nose}. The form of the various $\nabla_{\hbox{top}}^k$'s is given in
Tables~\tabref{heads} where the tops for all $k$ are included
for puposes of comparison. For $k=0$ the construction yields a space that
is best thought of as an elliptic fibration over $\IF_0 = \cp1\times\cp1$. For
$k=\half$ the construction yields a space that differs from $\cp4^{(1,1,1,6,9)}[18]$
owing to the presence of an extra constraint. In all other cases it is easy to see
that this construction gives again the space
$\cp4^{(1,1,2k,4k{+}4,6k{+}6)}[12k{+}12]$.
\topinsert
\vbox{\offinterlineskip
$$\vbox{\offinterlineskip\halign{
\strut # height 15pt depth 8pt&\quad$#$\quad\hfil\vrule
&\quad$#$\quad\hfil\vrule\cr
\noalign{\hrule}
\vrule&\hfil k&\hfil \nabla_{\hbox{top}}^k\cr
\noalign{\hrule\vskip3pt\hrule}
\vrule&0,\,\half,\,1  &\ca{T}^1\cr
\vrule&{3\over 2}     &\ca{T}^1\cup (0,1,1,2)\cr
\vrule&2              &\ca{T}^2\cr
\vrule&{5\over 2},\,3 &\ca{T}^3\cr
\vrule&{7\over 2},\,4 &\ca{T}^4\cup \{(0,1,0,0),\,(0,2,0,1),\,(0,3,1,2)\}\cr
\vrule&{9\over 2},\,5,\,{11\over 2},\,6 &\ca{T}^6\cr
\noalign{\hrule}
}}
$$
\tablecaption{heads}{The tops for each of the chains with
$k = 0,{1\over 2},\ldots,6$.}
}
\endinsert
The point of view that we adopt here is that the polyhedron picture is to be
preferred over the chains of projective spaces that were studied by
Aldazabal {it et al.\/}\cite{\rAFIQ}. We have already seen that points can sometimes
be added to the polyhedron without changing the Hodge numbers that were the basis for
the identifications and that when this is done the polyhedra can be seen to fall into
a regular sequence. This proves to be true of the tops also and we will return after
understanding the structure of the bottoms to cast the tops in their final form.

The polyhedra corresponding to restored gauge symmetries
are constructed, as explained previously, by leaving the tops
alone and modifying the bottoms according to Table~\tabref{bottoms}.
In this way we obtain Hodge numbers in agreement with what we expect
on the basis of the dual heterotic picture.
The Hodge numbers recorded in Tables~A\tabref{Hodge} and A\tabref{moreHodge} of
the appendix have been computed directly from the polyhedra and can be
seen to agree with eq.~\eqref{hodgeformulas} for $k=0,1/2,\cdots,6$.
\subsection{Non-perturbative effects}
We have seen that symmetry restoration in a terminal heterotic model can
be interpreted as adding points in the $\nab{4}_{\hbox{bot}}$
piece of $\nab{4}^k$, or equivalently, as imposing certain
additional conditions on the monomial deformations that determine
the dual Newton polyhedron. Notice that in this process, the $K3$
fibers are modified. We have then a qualitative explanation of our results
since, as explained by Aspinwall \cite{\rAsp}, the actual non-Abelian
structure of the group that is perturbatively visible is related
in turn to the structure of the $K3$ fibers. The arguments of
refs.~\cite{\rAL,\rAsp,\rAG} also provide some hints for how
to look for gauge groups that cannot be seen perturbatively. The
basic idea is to include the effect of degenerate fibers or to modify
the \cp1\ part of the fibration. Some of this can be done torically and
amounts to adding points outside the plane of the $K3$.

Motivated by the shape of $\nab{4}^{k,SU(1)}$, we have noticed that
adding points \hbox{$(1,2k{-}j,2,3)$}, $j=1, \cdots , 2k+2$ to this polyhedron
is always consistent with reflexivity. Moreover,
the Hodge numbers in this new sequence of polyhedra have a rather
interesting pattern as we now describe.
The transitions along the new branch are characterized by
$$\Delta h_{12} = -29 \quad\quad ; \quad\quad \Delta h_{11}=1 ~.
\eqlabel{hnp}$$
Hence, as implied by eq.~\eqref{newanom}, they can be explained as
transitions in which $n_T \to n_T + 1$. Since, eq.~\eqref{newinst} requires
$(d_1+d_2) \to (d_1+d_2 - 1)$, this corresponds to shrinking of an
instanton \cite{\rDMW, \rSW}. The result \eqref{hnp} is consistent
with $d_1 \to d_1 -1$ while $d_2$ is kept fixed. The first $E_7$ can
be completely broken for $d_1 \geq 10$ so that to arrive at $SU(1)$ the
original $d_1=12+2k$ can only be decreased in one unit $2+2k$
times as we have observed.

The new sequence of polyhedra has then a
non-perturbative interpretation. The observed transitions are
dual to some heterotic dynamics that can only be seen in $M$-theory.
The result \eqref{hnp} is also compatible with un-Higgsing of a
non-perturbative $SU(2)$ accompanied by exactly 16 doublets. When
$k=0$, eq.~\eqref{n2s} shows that un-Higgsing of a {\it perturbative}
$SU(2)$ precisely comes together with 16 doublets. Thus, in this case,
the same group structure can appear along the perturbative and
non-perturbative branches. Similar results were noticed in
ref.~\cite{\rDMW} for the heterotic vacuum and in ref.~\cite{\rAG}
for the type II vacuum. In our approach this is manifest given the
$\IZ_2$ symmetry $x_1 \leftrightarrow x_2$ of the $k=0$ polyhedron
\eqref{nose} that corresponds to exchange of the $\cp1$'s in
$\IF_0 = \cp1\times\cp1$. Thus, adding points $\widetilde{pt_i}^{(j)}$
obtained from $pt_i^{(j)}$ by $x_1 \leftrightarrow x_2$ leads to the same
perturbative groups but with a non-perturbative interpretation since
we have left fixed the generic $K3$~fiber.

We have also considered the effect of adding sequentially the points
$(1, 2k-j,2,3)$ for $j=1,2,\ldots,$ to the polyhedron $\nab{4}^{H,k}$
for all the $H$ that we have identified. For
$j\le j_H$, with $j_H$ an upper bound depending on $H$, this
always leads to a reflexive polyhedron and the Hodge numbers for these
sequences also have an interesting pattern.
To be more precise, we denote by $\nabla^{H,k,j}$ the augmented polyhedron:
$$
\nabla^{H,k,j} = \nab{4}^{H,k}\cup
\bigl\{ (1, 2k-1,2,3),~(1, 2k-2,2,3),~\ldots,~(1, 2k-j,2,3)\bigr\}$$
The extra points are being added to the two-face of $\nab{4}^{H,k}$
given by $x_3 = 2$ and $x_4 = 3$. Consider now the Hodge numbers
of $\nabla^{H,k,j}$ which, of course, depend on $(H,k,j)$.
If we fix $H$ and $k$ and consider the differences,
$(\D h_{11}, \D h_{21})$, between $\nabla^{H,k,j+1}$ and
$\nabla^{H,k,j}$ then these differences are
observed to be independent of $k$ and of
$j$ for $0\le j\le j_H$, and hence depend only on the group $H$.
Specifically, by an enumeration of cases, we find
 $$
\Delta h_{12} = -29 +\half b_H~,~~~\Delta h_{11}=1 ~,
\eqlabel{hnp2}$$
where the coefficients $b_H$ are those given in Table~\tabref{bottoms}.
Our interpretation is that the transitions $\nabla^{H,k,j} \to
\nabla^{H,k,j+1}$
correspond to processes in which
 $$
n_T \to n_T + 1~,~~~~d_1 \to d_1 - 1$$
and $d_2$ is kept fixed. The
term proportional to $b_H$ in $h_{12}$ arises because the decrease
in $d_1$ effectively implies $k \to k - \half$.
The bound $j_H$ depends on the values of $d_1$ that allow breaking
to $H$. This can be determined by looking at the number of fields
in the various representations and the breaking patterns. In this
way we obtain results that agree with the observed values of $j_H$ obtained
from
the polyhedra by continuing the sequence $\nabla^{H,k,j}$ until the polyhedra
cease to be reflexive.

Non-perturbative effects related to changes in the number of tensor
multiplets also provide an interpretation of the remaining reflexive
polyhedra obtained by systematically enlarging the bottoms as explained
in section 3.1. In all the cases in the lower block of Table~\tabref{bottoms}, the
Hodge numbers computed from the polyhedra take the form
$$\eqalign{
h_{11} &= {\rm rank}\, G_2^{(0)}(k) + {\rm rank}\,H +
\tilde{n}_T + 2\cropen{5pt}
h_{21} &= 272 + {\rm dim}\,G_2^{(0)}(k) + {\rm dim}\,H - 29\tilde{n}_T
- a_H - b_H \tilde{k}~.\cr} \eqlabel{hodgeformulasnp}$$
where $\tilde{n}_T$ is given in Table~\tabref{bottoms} and
$\tilde{k} = {1\over 2}(2k+ n_T(k) - \tilde{n}_T)$.
If we define $\tilde{d}_1=12+2\tilde{k}$,
the enlarged polyhedra then appear to
correspond to heterotic compactifications
with instanton numbers $(\tilde{d}_1, d_2)$, $\tilde{n}_T$ tensor
multiplets and an enhanced group $H\times G_2^{(0)}(k)$.
For example, when $H=E_8$ with $\tilde{k}=-6$ and
$k={9\over 2},\,5,\,{11\over 2},\,6$, the enhanced group is
$E_8\times E_8$, with $\tilde{d}_1=d_2=0$ and $\tilde{n}_T=25$.
The Hodge numbers obtained from \eqref{hodgeformulasnp},
$h_{11}=h_{21}=43$, are characteristic of a compactification
with 25 tensor multiplets\
\REF\rAIU{G.~Aldazabal, L.~E.~Ib\'a\~nez and A.~Uranga, unpublished.}
\cite{\rSW, \rAIU}. Note that $b_H=0$ in all these cases, reflecting
the fact that $\tilde{d}_1$ is independent of $\tilde{k}$ and hence,
so is the $H$ charged matter content.
\subsection{Irreducibility}
It is of interest to examine whether the $\nabla^{SU(1),k}$
polyhedra corresponding to the terminal
groups of each chain are irreducible \cite{\rACJM,\rCGGK}. It seems that this
should be so since it is not possible to break the symmetry further. Thus the
irreducibility of the polyhedra is a consistency check on duality and the
identity of the webs. By irreducible here we mean torically irreducible. That
is there is no reflexive sub-polyhedron of the given polyhedron. There is a
stricter notion of irreducibility which is more appropriate which is that the
given
manifold contains no (complex) curves that can be blown down without violating
the
condition $c_1 = 0$. Thus it is possible for a manifold to be reducible but to
be
torically irreducible. Of course if a manifold is irreducible in the strict
sense
it must also be torically irreducible. Thus the toric irreducibility of the
manifolds is a consistency check on duality though a weaker one than veryfying
irreducibility in the strict sense.

In virtue of the simple structure of the polyhedra the question  of the toric
irreducibility of the terminal manifolds of the chains can be examined quite
simply.  Consider the terminal model of the $k$'th chain. This has the points:
 $$\eqalign{
&\llap{*}\,( -1,\- 0,\- 2,\- 3)\cr
&\llap{*}\,(\- 0, -1,\- 2,\- 3)\cr
&\llap{*}\,(\- 0,\- 0, -1,\-  0)\cr
&\llap{*}\,(\- 0,\- 0,\- 0,  -1)\cr
&(\- 0,\- 0,\- 0,\- 0)\cr
&(\- 0,\- 0,\- 0,\- 1)\cr}
\hskip50pt
\eqalign{
&(\- 0,\- 0,\- 1,\- 1)\cr
&(\- 0,\- 0,\- 1,\- 2)\cr
&(\- 0,\- 0,\- 2,\- 3)\cr
&(\- 0,\- 1,\- 2,\- 3)\cropen{-2pt}
&\hskip20pt\vdots\hskip30pt\vdots\cropen{-2pt}
&\llap{*}\,(\- 1, 2k,\- 2,\- 3)~.\cr}
$$
In this list we have distinguished by asterisks five points which cannot be
discarded if we are to find a reflexive sub-polyhedron. The first point of the
list is the only one with $x_1 < 0$. This point cannot be omitted since then
the origin would lie in a face. This point is also a vertex. Similarly the last
point is the only one with $x_1>0$ and cannot therefore be omitted and is also a
vertex. The second, third and fourth points are the only ones that have negative
entries in the second, third and fourth columns respectively so these points must also
be retained and are also vertices. Since we have to retain these five points we must
retain also their convex hull. We can show now that this contains the tetrahedron
$\ca{T}^k$. The third and fourth points of $\nab{4}^k$ are vertices of $\nab{2}$. The
third vertex of~$\nab{2}$~is
 $$
(\- 0, \- 0, \- 2,\- 3) = {1\over 2k+2}\bigl\{
( -1,\- 0,\- 2,\- 3) + (\- 1, 2k,\- 2,\- 3) + 2k (\- 0,\- 1,\- 2,\- 3)
\bigr\}$$
and is therefore a point of the convex hull as is
 $$
(\- 0,\- k,\- 2,\- 3) = \half\bigl\{ ( -1,\- 0,\- 2,\- 3) +
(\- 1, 2k,\- 2,\- 3)\bigr\} $$
which is the top vertex of $\ca{T}^k$.

For some of the polyhedra the above observations already show
that the convex hull contains the entire polyhedron which is therefore
irreducible. For the remaining cases the only points that could perhaps be
omitted are subsets of $\nab{3}_{\hbox{top}}^k\setminus\ca{T}^k$. An
enumeration of possibilities, however, shows that all the polyhedra are
irreducible except for the single case of
the $k=\half$ polyhedron for which it is possible to omit the point $(0,1,2,3)$
which, in this case, is a vertex. The resulting polyhedron is the dual of the
Newton polyhedron of $\cp4^{(1,1,1,6,9)}[18]$. This manifold is not a
$K3$-fibration so for the case $k=\half$ the polyhedron is irreducible in the
weaker sense that it contains no reflexive subpolyhedron that is a
$K3$-fibration.

A final remark is perhaps in order. While there is no minimal polyhedron that
is contained in all the polyhedra that we have discussed it would be of
interest to know if there is a maximal polyhedron. We do not know if this
is the case. However if we denote by
${}^3\widehat{\nabla}^6_{\hbox{top}}$ the bottom formed by reflecting
${}^3\nabla^6_{\hbox{top}}$ in $\nab{2}$ (this is in fact the bottom
$\{ pt^{(6)}_1,\ldots,pt'_5 \}$, corresponding to the group $E_8$, from
Table~\tabref{bottoms}) then the
polyhedron
 $$\displaylines{
(-1,\- 0,\- 2,\- 3)~~~\crm
(\- 1,~ 12,\- 2,\- 3)~,~~(\- 1,~ 11,\- 2,\- 3)~,\ldots,~
(\- 1,-11,\- 2,\- 3)~,~~(\- 1,-12,\- 2,\- 3)\crm
{}^3\nabla^6_{\hbox{top}}~,~~{}^3\widehat{\nabla}^6_{\hbox{top}}\cr} $$
is a reflexive elliptic $K3$ fibration and contains all the polyhedra
that do not contain $pt'_6$ or~$pt'_7$.
\subsection{The final form of the polyhedra and the Dynkin diagrams}
In an earlier version of this paper the groups corresponding to the reflexive
polyhedra were identified on the basis of their hodge numbers. This led, in some
cases, to incorrect identifications and ambiguities owing to the fact that a
knowledge of the Hodge numbers alone does not always identify the group uniquely.
Thus manifolds corresponding to the groups $SO(8)$, $SO(9)$ and $F_4$, for example,
have the same Hodge numbers as do $SU(2){\times}SU(2)$ and $SO(5)$. These
ambiguities were resolved by Bershadsky {\it et al.\/}\
\Ref{\sixauthors}{M.~Bershadsky, K.~Intriligator, S.~Kachru, D.~R.~Morrison,
V.~Sadov and C.~Vafa, \npb{481} (1996) 215, hep-th/9605200.}\ 
who were able to identify the groups on the basis of the
singularity structure. Once the ambiguity is resolved however a very beautiful fact
emerges which is that the Dynkin diagram of the group and also the extended diagram
can be read off from the $K3$-polyhedron. To see this it is instructive to 
recall that the Hodge numbers $(h^{11},h^{21})$ of
a hypersurface $\M$ of this family may be calculated directly in terms of data
derived from the Newton
polyhedron. Let $\hbox{pts}(\D)$ denote the number of integral points of
$\D$ and let $\D_r$ denote the set of $r$-dimensional faces of $\D$. Write also
$\hbox{int}(\th)$ for the number of integral points interior to a face, $\th$,
of $\D$ and define similar quantities with $\D$ and $\nabla$ interchanged.
Duality provides a unique correspondence between an $r$-dimensional face,
$\th$, of $\D$ and a $(3-r)$-dimensional face $\tilde\th$ of $\nabla$. With
this notation the formulae
\REFS\rbatmir{V. Batyrev, J.\ Alg.\ Geom. {\bf 3} (1994) 493,
alg-geom/9310003.}
\REFSCON\rmondiv{P. S. Aspinwall, B. R. Greene and D. R. Morrison, Int.\
Math.\ Res.\ Notices (1993) 319, alg-geom/9309007.}
\refsend\
for the Hodge numbers are
 $$\eqalign{
h^{21}(\D) &= \hbox{pts}(\D) - \sum_{\th\in\D_3} \hbox{int}(\th) +
\sum_{\th\in\D_2} \hbox{int}(\th)\,\hbox{int}(\tilde\th) - 5~,\cropen{5pt}
h^{11}(\D) &= \hbox{pts}(\nabla) -
\sum_{\tilde\th\in\nabla_3} \hbox{int}(\tilde\th) +
\sum_{\tilde\th\in\nabla_2} \hbox{int}(\tilde\th)\,\hbox{int}(\th) - 5~.\cr}
\eqlabel{batyrev}$$
The degrees of freedom to redefine the coordinates of the embedding space is
accounted for by the points interior to codimension-one faces of the polyhedra. This
accounts for the fact that these points do not contribute to the Hodge numbers. The
terms
$$\sum_{\th\in\D_2} \hbox{int}(\th)\,\hbox{int}(\tilde\th)~~~~\hbox{and}~~~~
\sum_{\tilde\th\in\nabla_2} \hbox{int}(\tilde\th)\,\hbox{int}(\th)$$
account for non-toric deformations of the manifold.

We present in the Appendix the figures for each of the bottoms of
Tables~\tabref{bottoms} and \tabref{bottomstwo}. In the electronic version of this
article the figures are in colour and the integral points of the polyhedron are
coloured according to how they contribute to Batyrev's formulas
\eqref{batyrev}. The Dynkin diagrams are formed by the red points and lines. The
points are coloured according to the following rules:

\item{}{\sl --Black points: }These are the points that are interior to codimension
one faces of the 4D polyhedron and which are subtracted in Batyrev's formulas owing
to the fact that they correspond to the freedom to redefine the homogeneous
coordinates of the embedding space.

\item{}{\sl --Green points: }These are four of the five points that are subtracted
off in Batyrev's formulas. They consist of the three vertices of the triangle
corresponding to the torus and the first point connected to the point $(0,0,2,3)$ by
an edge. The fifth point is the first point connected to the point $(0,0,2,3)$ by
an edge in the top. When the polyhedron contains the point $pt'_1$ and its reflection
in the triangle these are the points of the $K3$ polyhedron corresponding to the
trivial group.

\item{}{\sl --Blue points: }These occur only ocaisionally as in the $E_8$ or $SO(13)$
polyhedra. These are the points that lie interior to codimension one faces of the
$K3$ polyhedron but are non-trivial owing to the fact that they {\it do not\/} lie in
codimension one faces of the 4D polyhedron. These points contribute to the number of
tensor multiplets. Finally we are left with the points that contribute to the rank of
the group

\item{}{\sl --Red points: }These give the Dynkin diagram of the group and, on
including the first point that is connected to the point $(0,0,2,3)$ by an edge we
see also the extended diagram of the group with the green point as the extending
point.

\noindent Of course we do not see the multiple lines of the Dynkin diagrams in the
edges of the polyhedron so what we see is the `skeleton' of the diagrams with all
lines given as single lines and no distinction made between long and short roots.
This being so it is fortunate that we see both the Dynkin diagram and the extended
diagram in each bottom since, apart from $SO(5)$ and $G_2$ for which the skeletons
of the diagrams are the same, these serve to uniquely identify the groups. For the
cases of $SO(5)$ and $G_2$ the groups are distinguished on the basis of the hodge
numbers.

For the bottoms of \tabref{bottomstwo} the result of omitting the point $pt'_1$ is
that now some of the points interior to the edges of the triangle of the torus now
contribute to the rank of the group since they are no longer interior to
codimension-one faces of the 4D-polyhedron. 

Figures \figref{arrowsa} and \figref{arrowsb} show the nesting of the bottoms with the
arrows denoting inclusion. Not all inclusions are shown; there are for example a
number of connections between the two diagrams. It is gratifying that the majority of
cases are understood as the straightforward inclusion of groups. Other cases such as
$E_{6b}\to SU_{6c}\to SU_{6b}$ would seem to indicate the possibility of
non-perturbative transitions.
\pageinsert
\figbox{\leavevmode\epsfxsize=5truein\vbox{\vskip.3in\hbox{
\hskip.5in\epsfbox{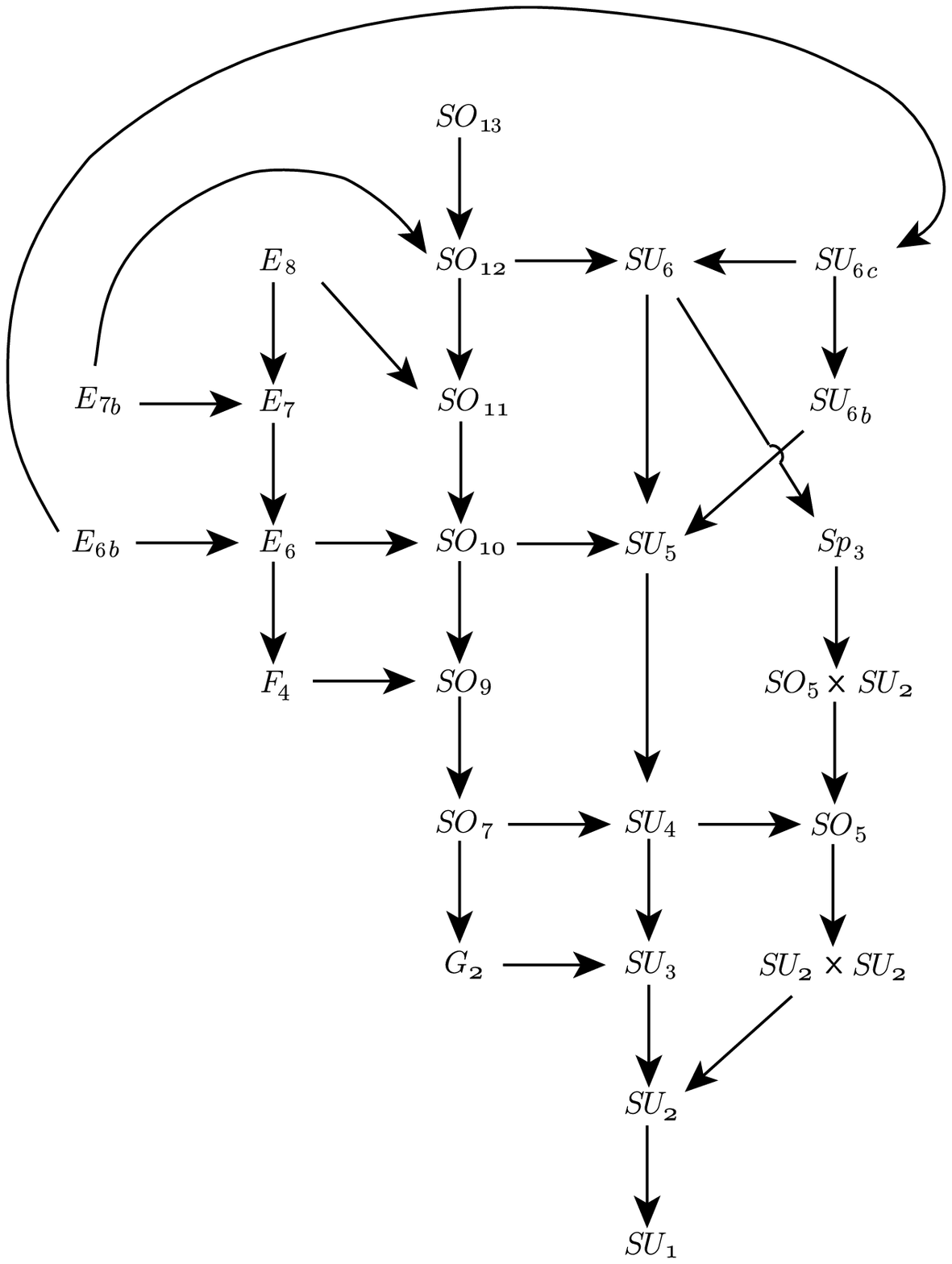}}}\vskip.2in}{\figlabel{arrowsa}}{Inclusion of
polyhedra for the polyhedra that contain the point $pt'_1$. The arrows denote
inclusion of the bottoms.}
\vfil
\endinsert
\pageinsert
\figbox{\leavevmode\epsfxsize=5truein\vbox{\vskip.4in\hbox{
\hskip.5in\epsfbox{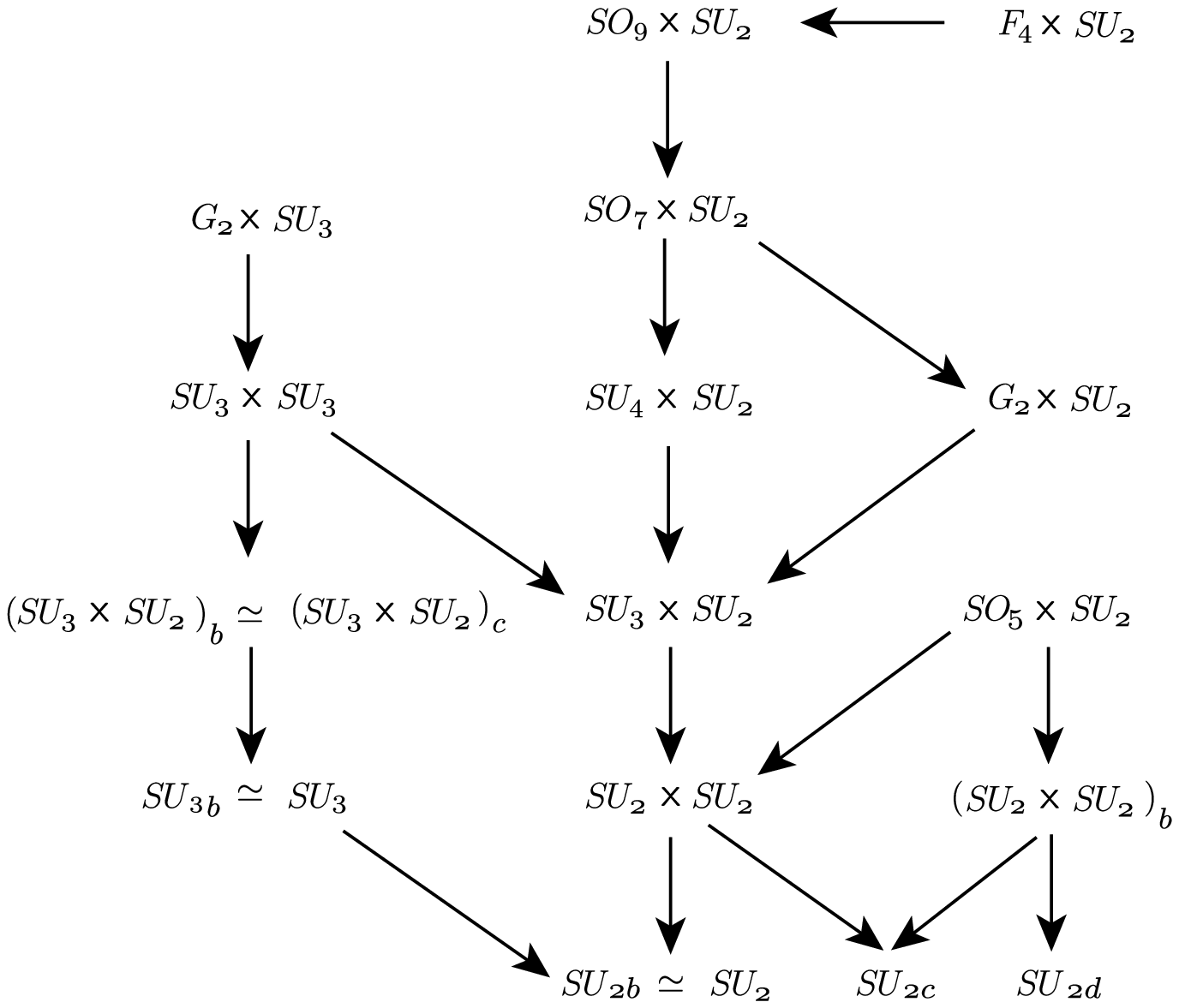}}}\vskip-.7in}{\figlabel{arrowsb}}{Inclusion of
polyhedra for the polyhedra that do not contain the point $pt'_1$. This figure has
been drawn on the assumption that some of the polyhedra correspond to the same \cym.
This hypothesized equivalence is denoted by $\asymp$ in the diagram.}
\vfil
\endinsert 
Finally we return to discuss the structure of the tops. As given in
Table~\tabref{heads}, these do not as pleasing an interpretation as the bottoms. In
particular not all of the points of the Dynkin diagrams are visible. For example the
top for $k=3/2$ corresponds to the group $G_2^{(0)} = SU(3)$ but is the reflection,
in the triangle of the torus, of the $SU(2)$ bottom. Additionally the $k=5/2$ and
$k=3$ tops are the same, being the reflection of the $F_4$ bottom, though the
corresponding groups are $G_2^{(0)} = F_4$ and $G_2^{(0)} = E_6$. This suggests that
the `missing' points should be added to the tops. This can be done by replacing the
tops by the reflections of the bottoms for the groups $G_2^{(0)}$ except that $SO(9)$
should be used instead of $SO(8)$ and $E_7$ in place of what we have called $E_7^{-}$.
The Hodge numbers are unchanged by this process. The reason that the points that
contribute to the rank of the group were not previously visible is that although the
previously given tops give the correct values for the Hodge numbers they correspond,
for those cases for which the Dynkin diagram is not fully visible, to manifolds with
non-toric parameters \ie to polyhedra for which the correction term in Batyrev's
formulas is non-zero. By adding the extra points to the tops the correction term
disappears so that now all the parameters are torically expressed.
\newpage
\section{fin}{Discussion}
In this article we have considered $K3\times T^2$,
$E_8\times E_8$ compactifications with instanton numbers $(d_1,d_2)$
and $n_T$ tensor multiplets. Matching of spectra \cite{\rKV,\rAFIQ},
and arguments based on $F$-theory \cite{\rVI,\rMV},
indicate that at the points of maximal symmetry
breaking heterotic theories of this type are dual to type IIA
compactifications on \cys\ that admit a $K3$ as well as an
elliptic fibration. The new contribution here is the observation that
sequences of reflexive polyhedra associated to these spaces are nested
in such a way as to reflect heterotic perturbative and non-perturbative
processes. This qualitative observation is supported by quantitative
agreement of the computed Hodge numbers and the number of tensor, vector
and hypermultiplets in the heterotic side. It leads also to the observation that the
Dynkin diagrams for the groups may be read off from the polyhedra.

The results displayed in Tables~\tabref{bottoms} and~~\tabref{bottomstwo} 
correspond to reflexive polyhedra that can be formed
by adding the points $pt_r^{(j)}$ to $\nabla_{\hbox{bot}}^{SU(1)}$.
It is also possible to extend the bottoms by adding points that are not
directly below the $pt_r$. It is clear that there are other groups that can be
realised in this way but a systematic investigation will require an efficient way of
enumerating possibilities. Also of considerable interest is the issue of which points
may be added in the fourth dimension. These four dimensional points are very
interesting since they are associated with non-perturbative effects.
We have touched on some of these issues with the polyhedra
$\nabla^{H,k,j}$ but this is clearly just scratching the surface of possibilities.

In the heterotic picture, $E_8 \times E_8$ is broken to a
generic $G_1 \times G_2$ group by background fields with
$(d_1,d_2)$ instanton numbers. In terms of polyhedra we have
only found the equivalent description of processes in which
$G_2$ remains maximally broken while $A$, $D$, and $E$ group
factors in $G_1$ are perturbatively restored. In the
non-perturbative effects that were observed, $G_2$ also remains
stable. Since we have not seen signals of $G_2$ dynamics,
expected at least for some $d_2$, we presume that they will arise
upon more generic manipulations of the polyhedra.

In summary, we have uncovered the beginnings of a dictionary that
translates between vector bundles and polyhedra. The precise nature
of this dictionary remains for future work.

While this article was being completed we received two articles\
\REFS\rWitIII{E.~Witten, hep-th/9603150.}
\REFSCON\rMVII{D.~R.~Morrison and C.~Vafa, hep-th/9603161.}
\refsend\
which overlap with the present work.

\vskip.5in
\vbox{
\acknowledgements
We had the benefit of useful conversations with G.~Aldazabal, P.~Aspinwall,
X.~de la Ossa, L. Ib\'a\~nez, V.~Kaplunovsky, S.~Katz, F.~Quevedo and
A.~Uranga.
This work was supported in part by the Robert A. Welch Foundation and
the NSF grant PHY-9511632. A.F. acknowledges a research grant Conicit-S1-2700
and a CDCH-UCV sabbatical fellowship.}
\vskip.5in
\chapno=-1
\section{appendix}{Appendix: Tables of Hodge Numbers and Figures}
Tables of Hodge numbers calculated from the polyhedra
are shown on the following pages together with
plots of a selection of tops and bottoms of the
polyhedra.
\newpage
$$
\def\skip{\hskip3.4pt}
\vbox{\offinterlineskip\halign{
\strut # height 15pt depth 8pt&\hfil\hskip5pt $#$ \skip \hfil\vrule
&\hskip10pt  \eightrm # \skip \hfil &\skip  \eightrm # \skip \hfil
&\skip  \eightrm # \skip \hfil &\skip  \eightrm # \skip \hfil
&\skip  \eightrm # \skip \hfil &\skip  \eightrm # \skip\hfil
&\skip  \eightrm # \skip\hfil&\skip  \eightrm # \hskip5pt\hfil\vrule
\cr
\noalign{\hrule}
\vrule&k&\hfil $SU(1)$&\hfil $SU(2)$&\hfil $SU(3)$&\hfil $SU(4)$&\hfil $SU(5)$
&\hfil $SO(10)$&\hfil $E_6$&\hfil $E_7$\cr
\noalign{\hrule\vskip3pt\hrule}
\vrule&0             &(243,  3)  &(214,  4)  &(197,  5)  &(182,  6)
&(167,  7)  &(164, 8) &(159, 9) &(152, 10)\cr
\vrule&{1\over 2}    &(243,  3)  &(202,  4)  &(179,  5)  &(160,  6)
&(142,  7)  &(138, 8) &(132, 9) &(124, 10)\cr
\vrule&1             &(243,  3)  &(190,  4)  &(161,  5)  &(138,  6)
&(117,  7)  &(112, 8) &(105, 9) &(96, 10)\cr
\vrule&{3\over 2}    &(251,  5)  &(186,  6)  &(151,  7)  &(124,  8)
&(100,  9)  &(94, 10) &(86, 11) &(76, 12)\cr
\vrule&2             &(271,  7)  &(194,  8)  &(153,  9)  &(122,  10)
&(95,  11)  &(88, 12) &(79, 13) &(68, 14)\cr
\vrule&{5\over 2}    &(295,  7)  &(206,  8)  &(159,  9)  &(124,  10)
&(94,  11)  &(86, 12) &(76, 13) &(64, 14)\cr
\vrule&3             &(321,  9)  &(220,  10) &(167,  11) &(128,  12)
&(95,  13)  &(86, 14) &(75, 15) &(62, 16)\cr
\vrule&{7\over 2}    &(348,  10) &(235,  11) &(176,  12) &(133,  13)
&(97,  14)  &(87, 15) &(75, 16) &(61, 17)\cr
\vrule&4             &(376,  10) &(251,  11) &(186,  12) &(139,  13)
&(100,  14) &(89, 15) &(76, 16) &(61, 17)\cr
\vrule&{9\over 2}    &(404,  14) &(267,  15) &(196,  16) &(145,  17)
&(103,  18) &(91, 19) &(77, 20) &(61, 21)\cr
\vrule&5             &(433,  13) &(284,  14) &(207,  15) &(152,  16)
&(107,  17) &(94, 18) &(79, 19) &(62, 20)\cr
\vrule&{{11}\over 2} &(462,  12) &(301,  13) &(218,  14) &(159,  15)
&(111,  16) &(97, 17) &(81, 18) &(63, 19)\cr
\omit{\vrule height 15pt depth10pt}
      &6             &(491,  11) &(318,  12) &(229,  13) &(166,  14)
&(115,  15) &(100, 16) &(83, 17) &(64, 18)\cr
\noalign{\hrule}
}}
$$
\nobreak\tablecaption{Hodge}{The Hodge numbers $(h^{21},  h^{11})$ calculated
from the polyhedra $\nabla^{k,  H}$ for the chain
$H=SU(1),  \ldots,  SU(5),\, SO(10),\, E_6,\, E_7$.}
\newpage
$$
\def\skip{\hskip4pt}
\vbox{\offinterlineskip\halign{
\strut # height 15pt depth 8pt&\hfil\hskip5pt $#$ \skip \hfil\vrule
&\hskip10pt  \eightrm # \skip \hfil &\skip  \eightrm # \skip \hfil
&\skip  \eightrm # \skip \hfil &\skip  \eightrm # \skip \hfil
&\skip  \eightrm # \skip \hfil &\skip  \eightrm # \hskip5pt\hfil\vrule
\cr
\noalign{\hrule}
\vrule&k&$SU(2)^2$ &$SU(2)^3$ &$SO(8)$~~~ &$SU(6)$ &$SU(6)_b$ &$SO(12)$\cr
\noalign{\hrule\vskip3pt\hrule}
\vrule&0             &(185,  5)  &(156,  6) &(175, 7) &(152, 8)
&(150, 8) &(149, 9)\cr
\vrule&{1\over 2}    &(165,  5)  &(132,  6) &(151, 7) &(125, 8)
&(122, 8) &(121, 9)\cr
\vrule&1             &(145,  5)  &(108,  6) &(127, 7) &(98, 8)
&(94, 8) &(93, 9) \cr
\vrule&{3\over 2}    &(133,  7)  &(92,  8)  &(111, 9) &(79, 10)
&(74, 10) &(73, 11)\cr
\vrule&2             &(133,  9)  &(88,  10) &(107, 11)&(72, 12)
&(66, 12) &(65, 13)\cr
\vrule&{5\over 2}    &(137,  9)  &(88,  10) &(107, 11)&(69, 12)
&(62, 12) &(61, 13)\cr
\vrule&3             &(143,  11) &(90,  12) &(109, 13)&(68, 14)
&(60, 14) &(59, 15)\cr
\vrule&{7\over 2}    &(150,  12) &(93,  13) &(112, 14)&(68, 15)
&(59, 15) &(58, 16)\cr
\vrule&4             &(158,  12) &(97,  13) &(116, 14)&(69, 15)
&(59, 15) &(58, 16)\cr
\vrule&{9\over 2}    &(166,  16) &(101,  17)&(120, 18)&(70, 19)
&(59, 19) &(58, 20)\cr
\vrule&5             &(175,  15) &(106,  16)&(125, 17)&(72, 18)
&(60, 18) &(59, 19)\cr
\vrule&{11\over 2}   &(184,  14) &(111,  15)&(130, 16)&(74, 17)
&(61, 17) &(60, 18)\cr
\omit{\vrule height 15pt depth10pt}
      &6             &(193,  13) &(116,  14)&(135, 15)&(76, 16)
&(62, 16) &(61, 17)\cr
\noalign{\hrule}
}}
$$
\nobreak\tablecaption{moreHodge}{The Hodge numbers $(h^{21},  h^{11})$
calculated from the polyhedra $\nabla^{k,  H}$ for the groups
$SU(2)^2$, $SU(2)^3$, $SO(8)$, $SU(6)$, $SU(6)_b$ and $SO(12)$.}
\newpage
\def\simplefig#1{\figbox{\leavevmode\epsfxsize=5truein\hbox{
\hskip.75in\epsfbox{simple#1.eps}}}{\figlabel{fig#1}}{Two views of the
polyhedra for each of the indicated groups.}\newpage}
\simplefig{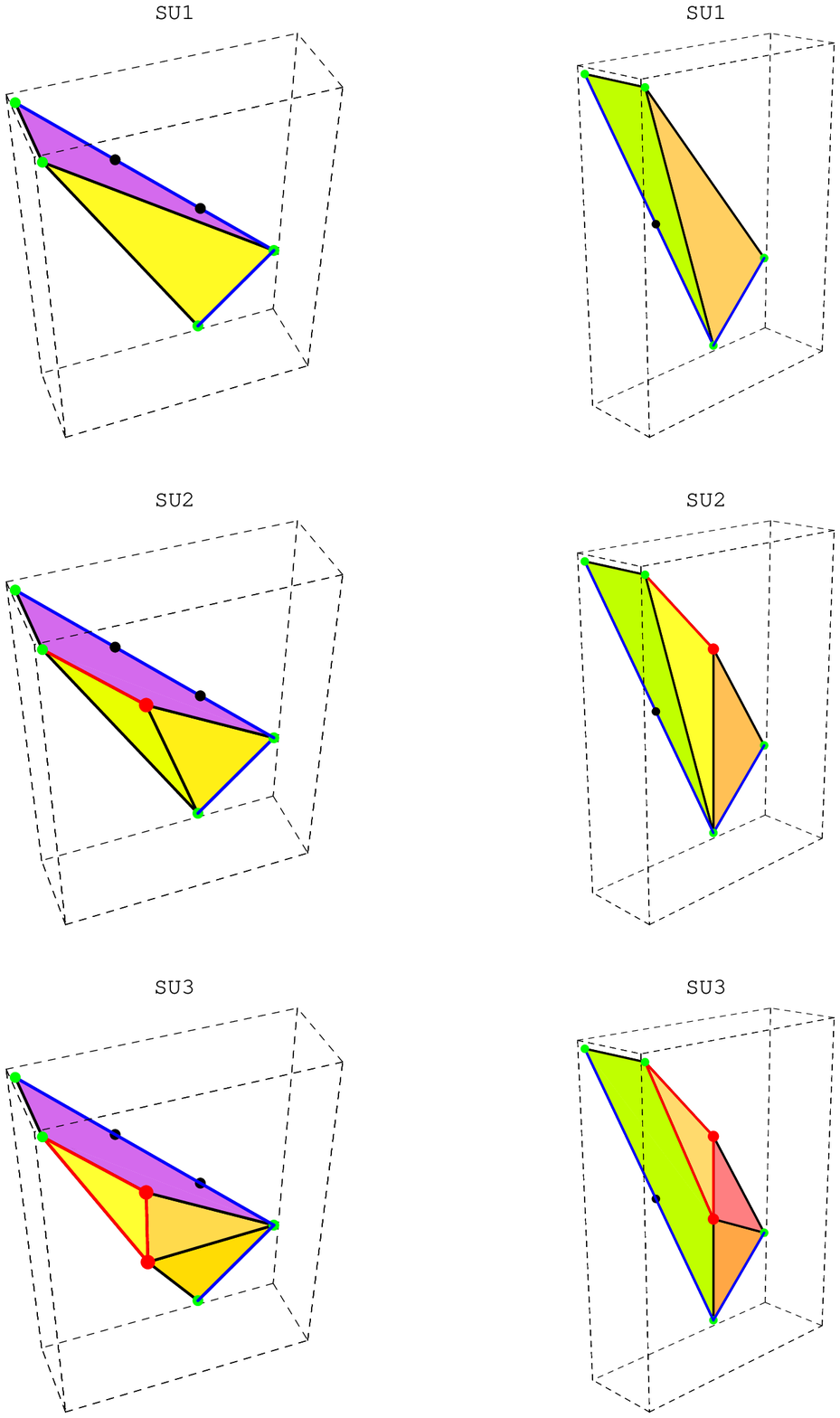}
\simplefig{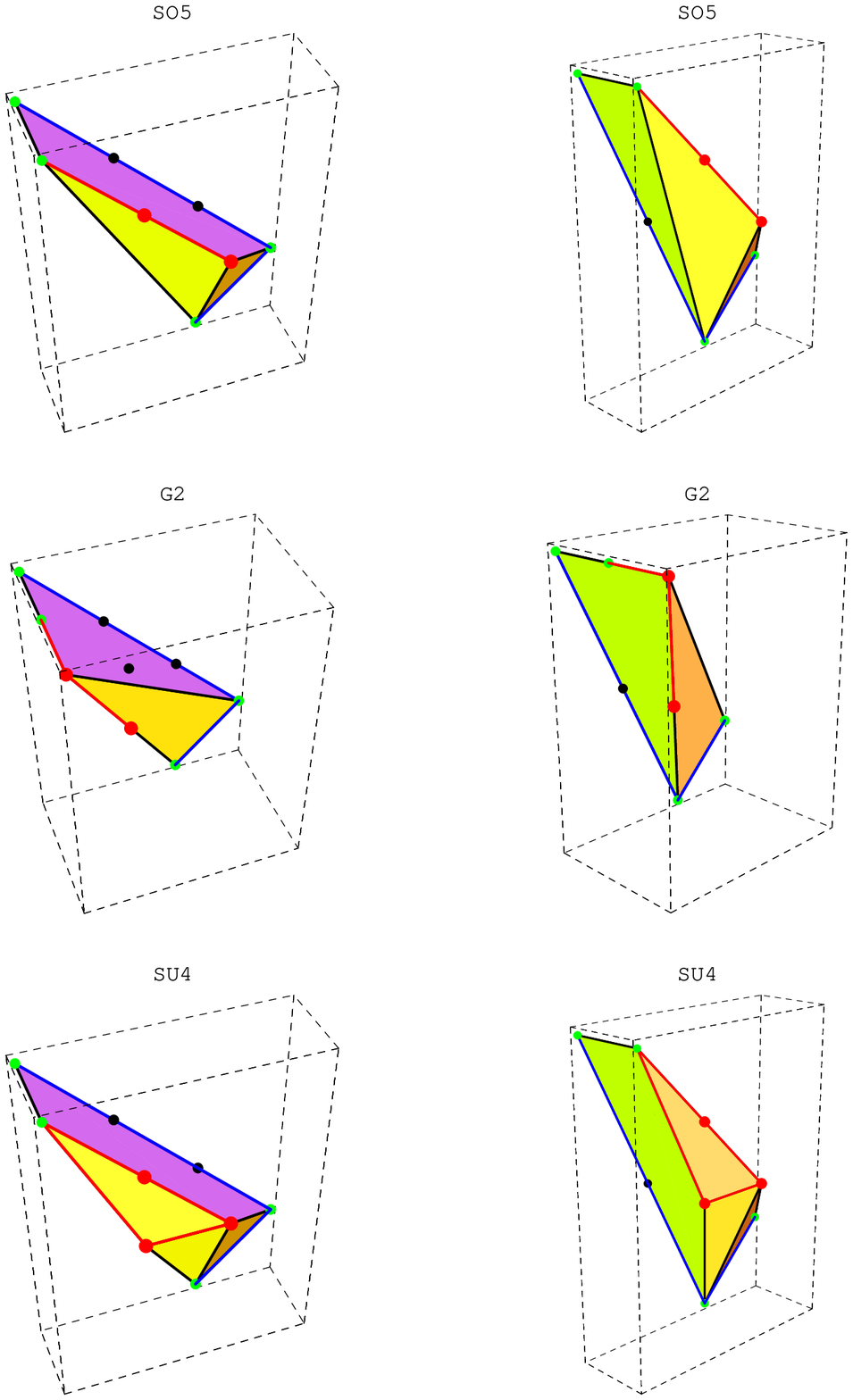}
\simplefig{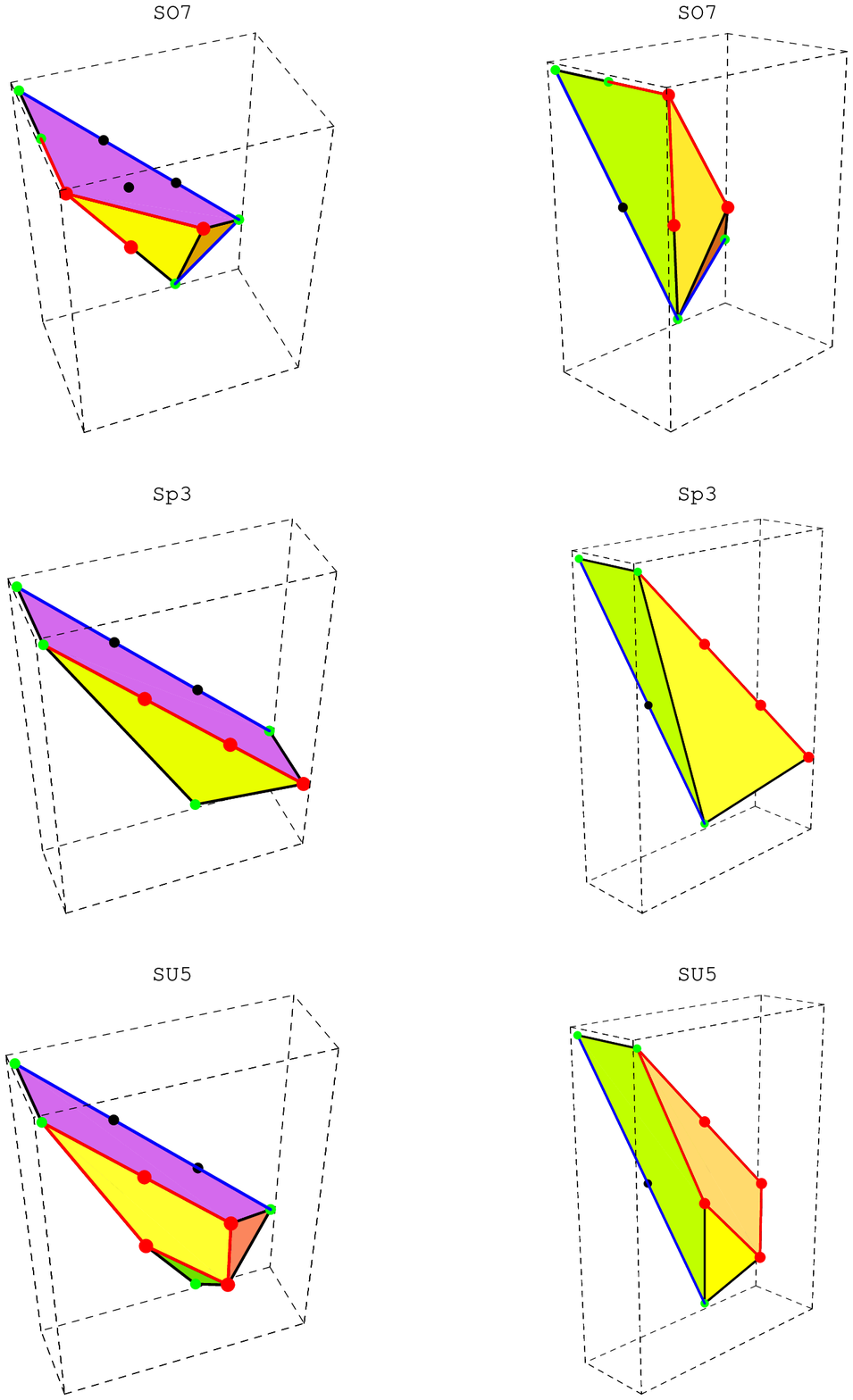}
\simplefig{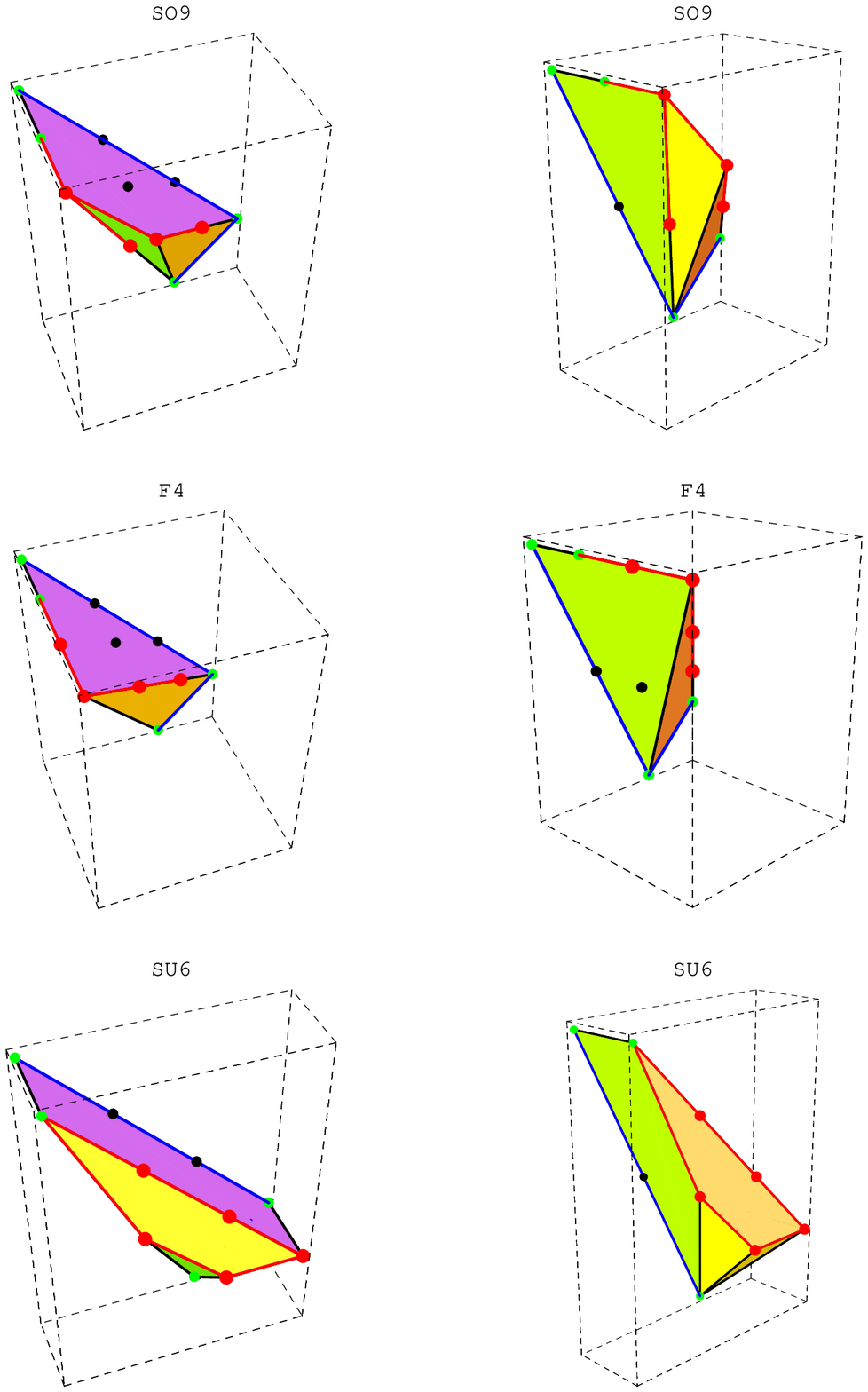}
\simplefig{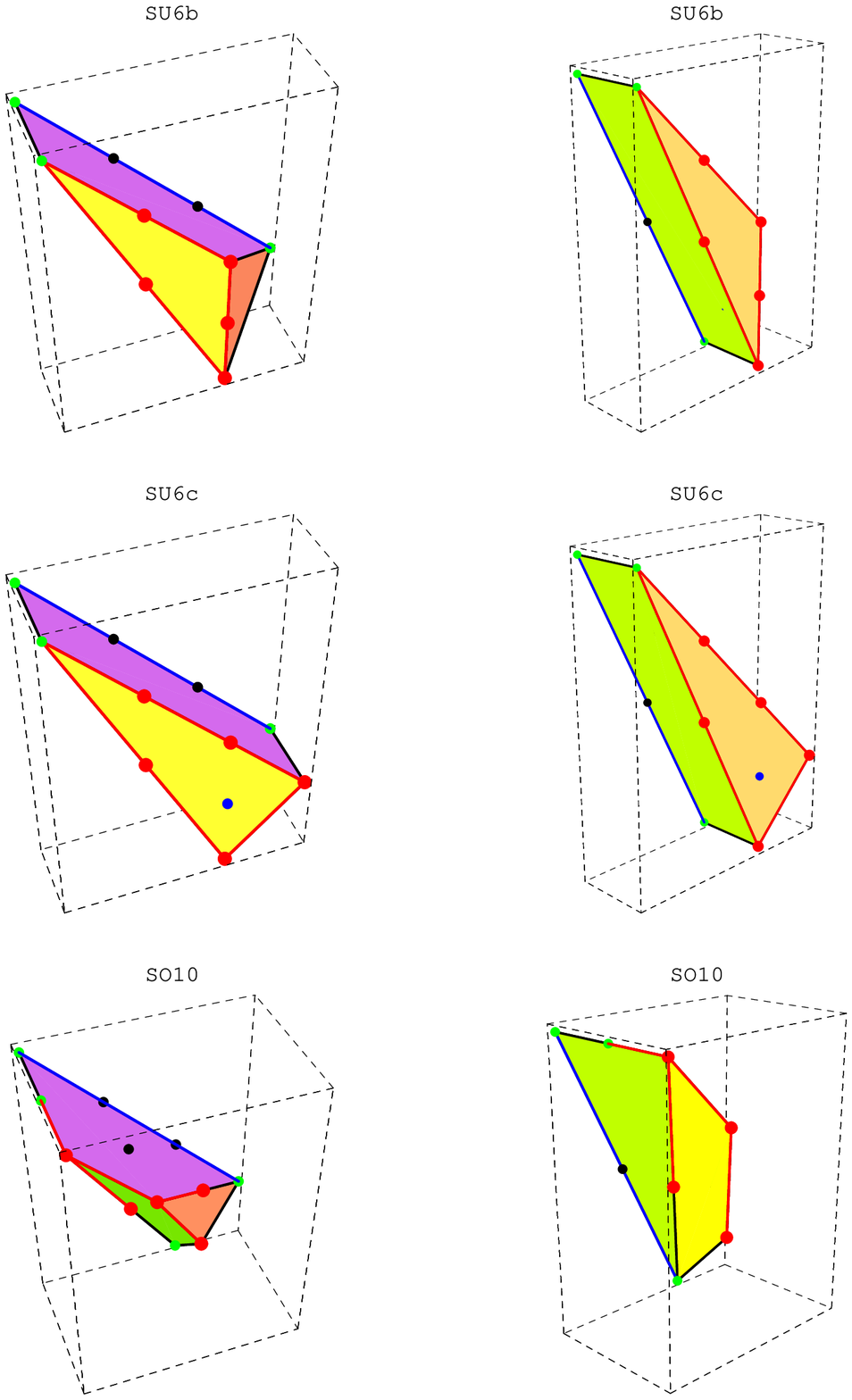}
\simplefig{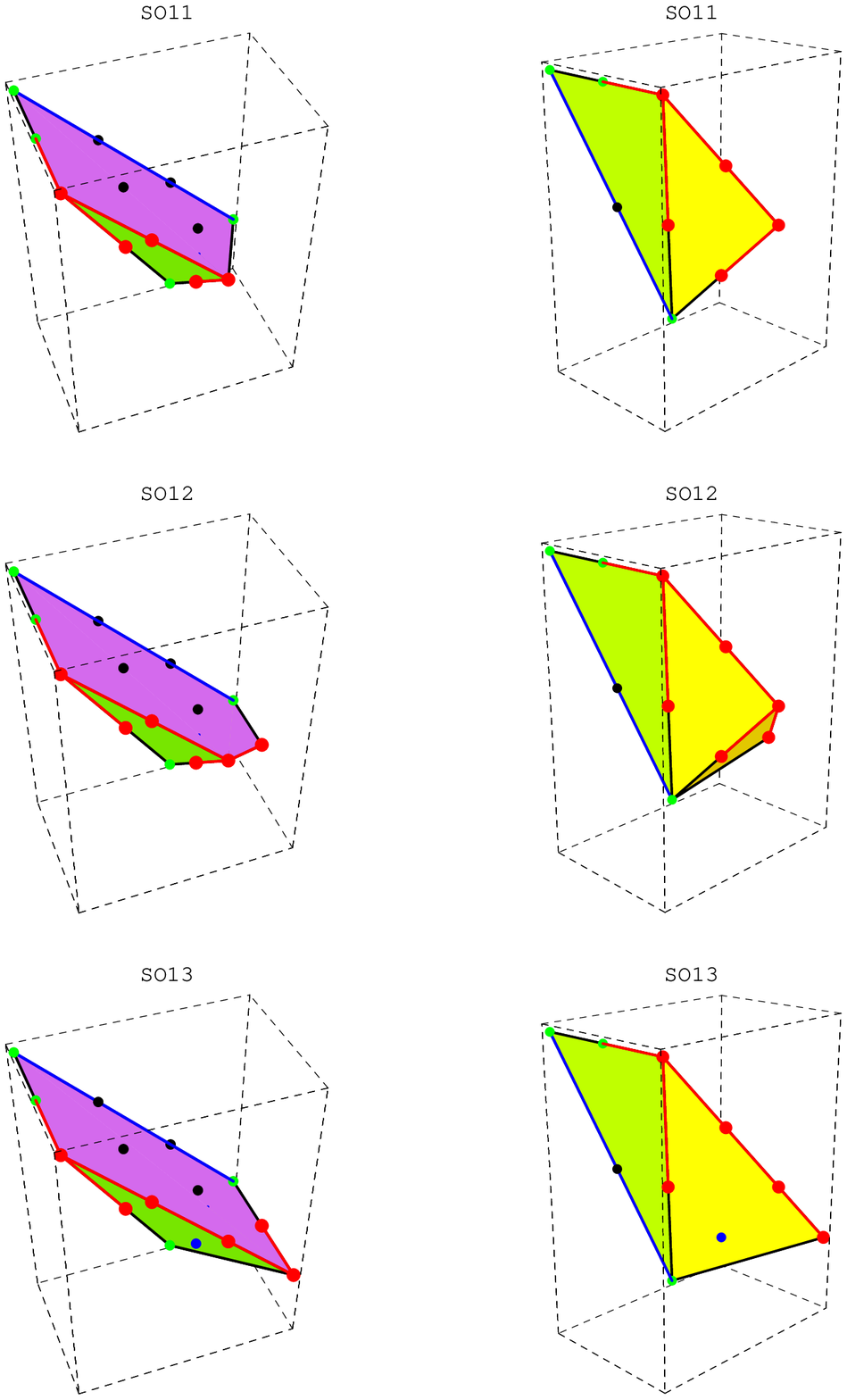}
\simplefig{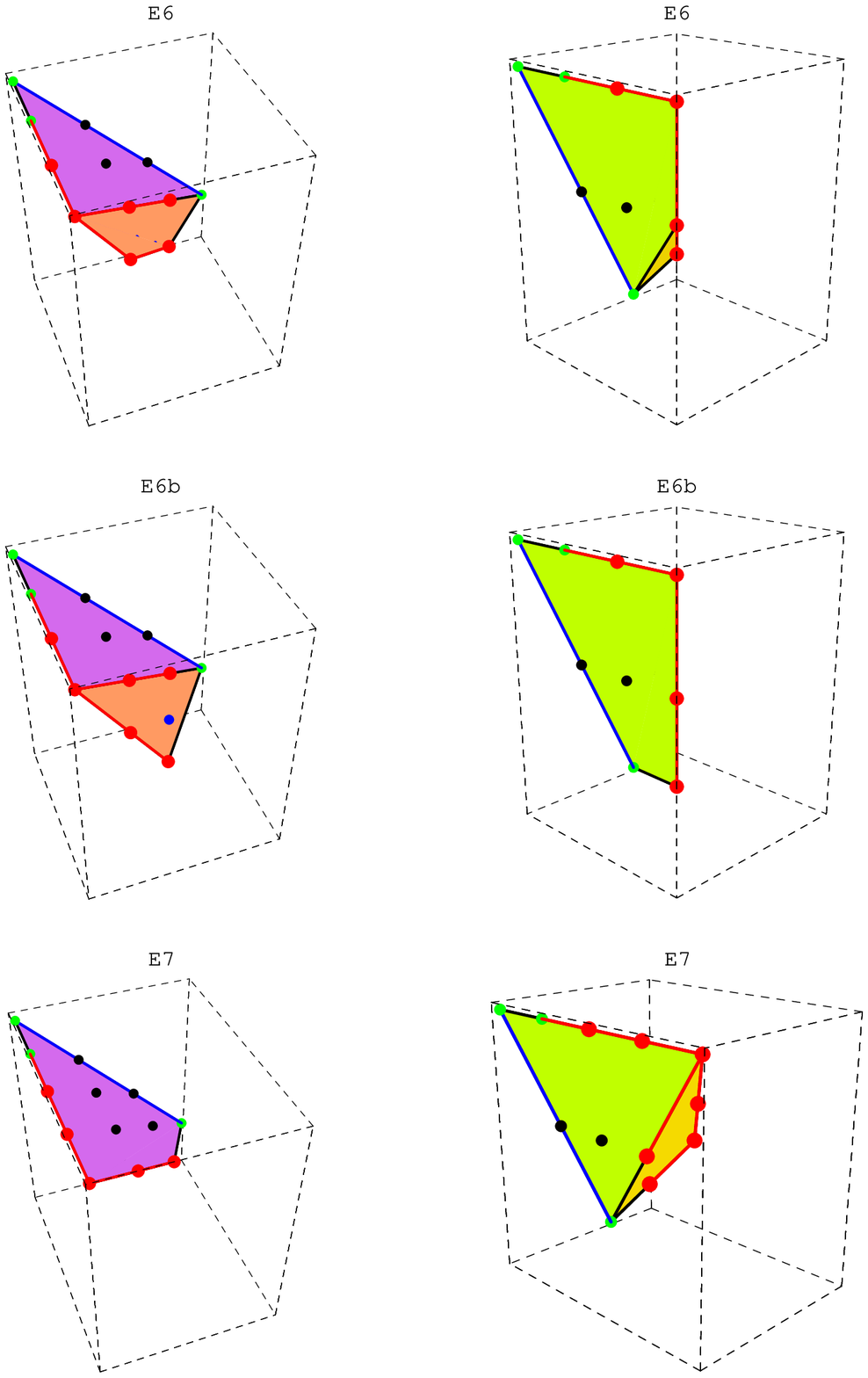}
\simplefig{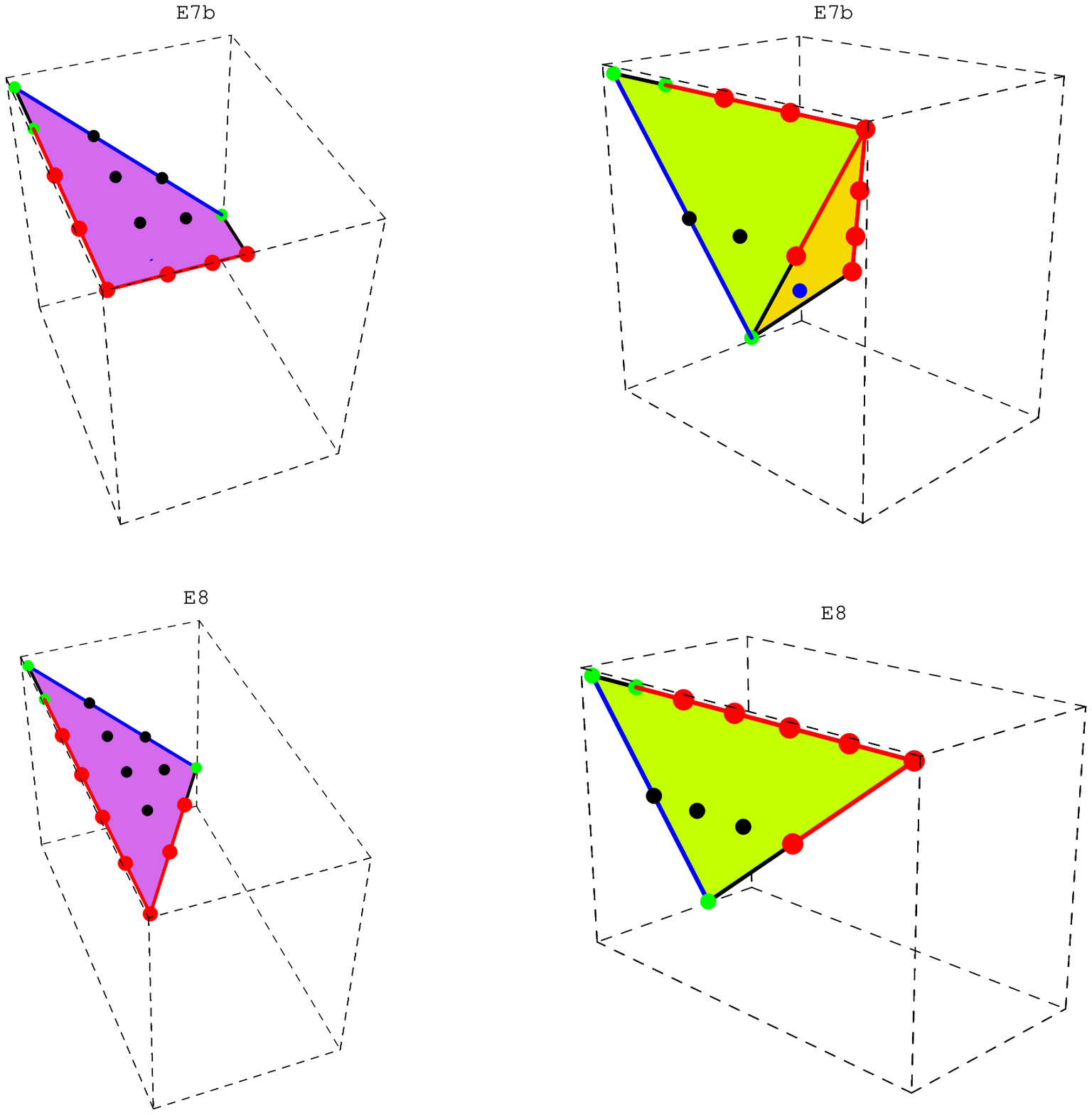}
\def\nonsimplefig#1{\figbox{\leavevmode\epsfxsize=5truein\hbox{
\hskip.75in\epsfbox{nonsimple#1.eps}}}{\figlabel{fig#1}}{Two views of the
polyhedra for each of the indicated groups.}}
 %
\nonsimplefig{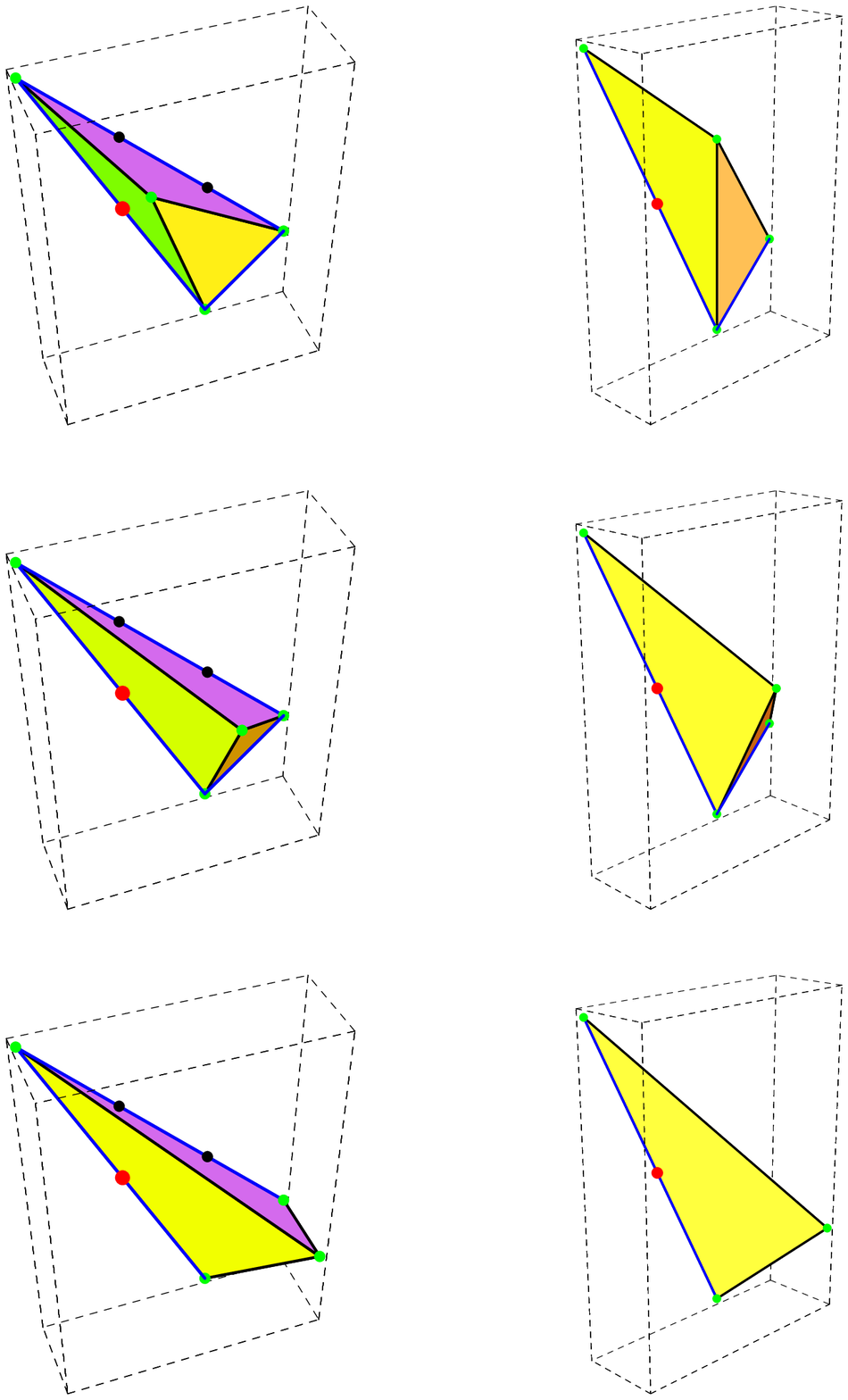}
\place{.3}{6.75}{$SU_{2b}$}
\place{.3}{4.375}{$SU_{2c}$}
\place{.3}{2}{$SU_{2d}$}
\newpage
\nonsimplefig{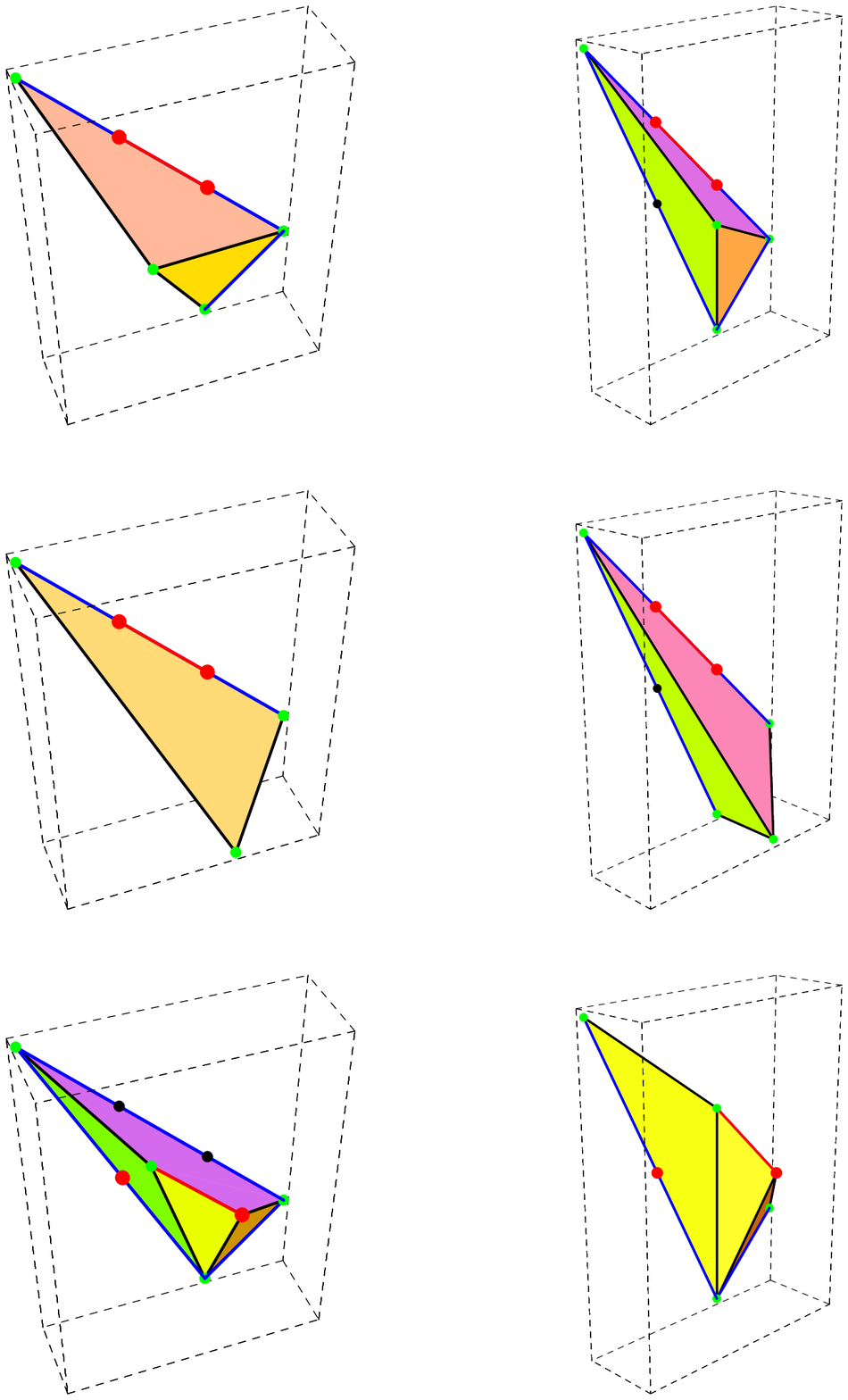}
\place{.3}{6.75}{$SU_{3b}$}
\place{.3}{4.375}{$SU_{3c}$}
\place{.3}{2}{$SU_2{\times}SU_2$}
\newpage
\nonsimplefig{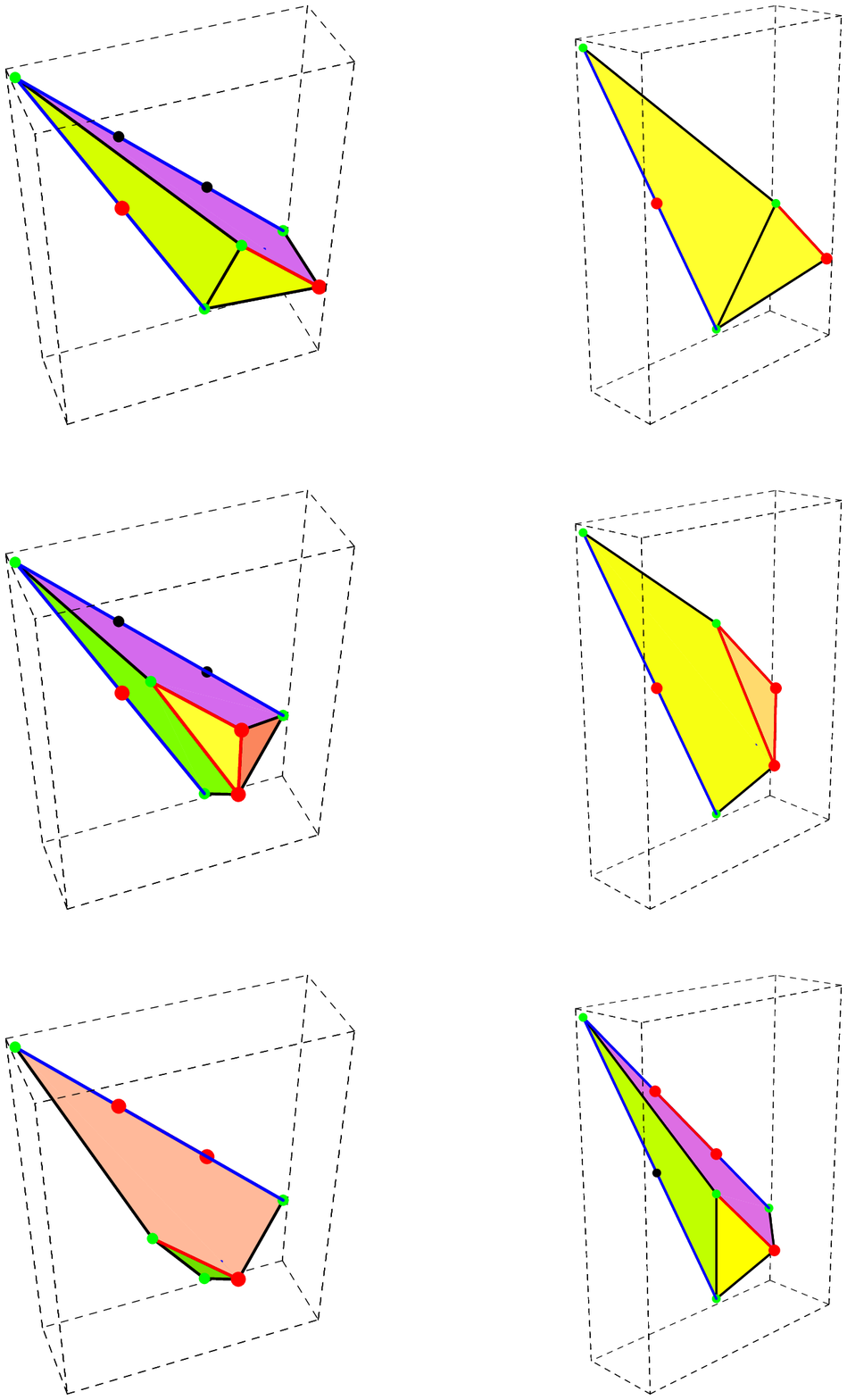}
\place{.3}{6.75}{$(SU_2{\times}SU_2)_b$}
\place{.3}{4.375}{$SU_2{\times}SU_3$}
\place{.3}{2}{$(SU_2{\times}SU_3)_b$}
\newpage
\nonsimplefig{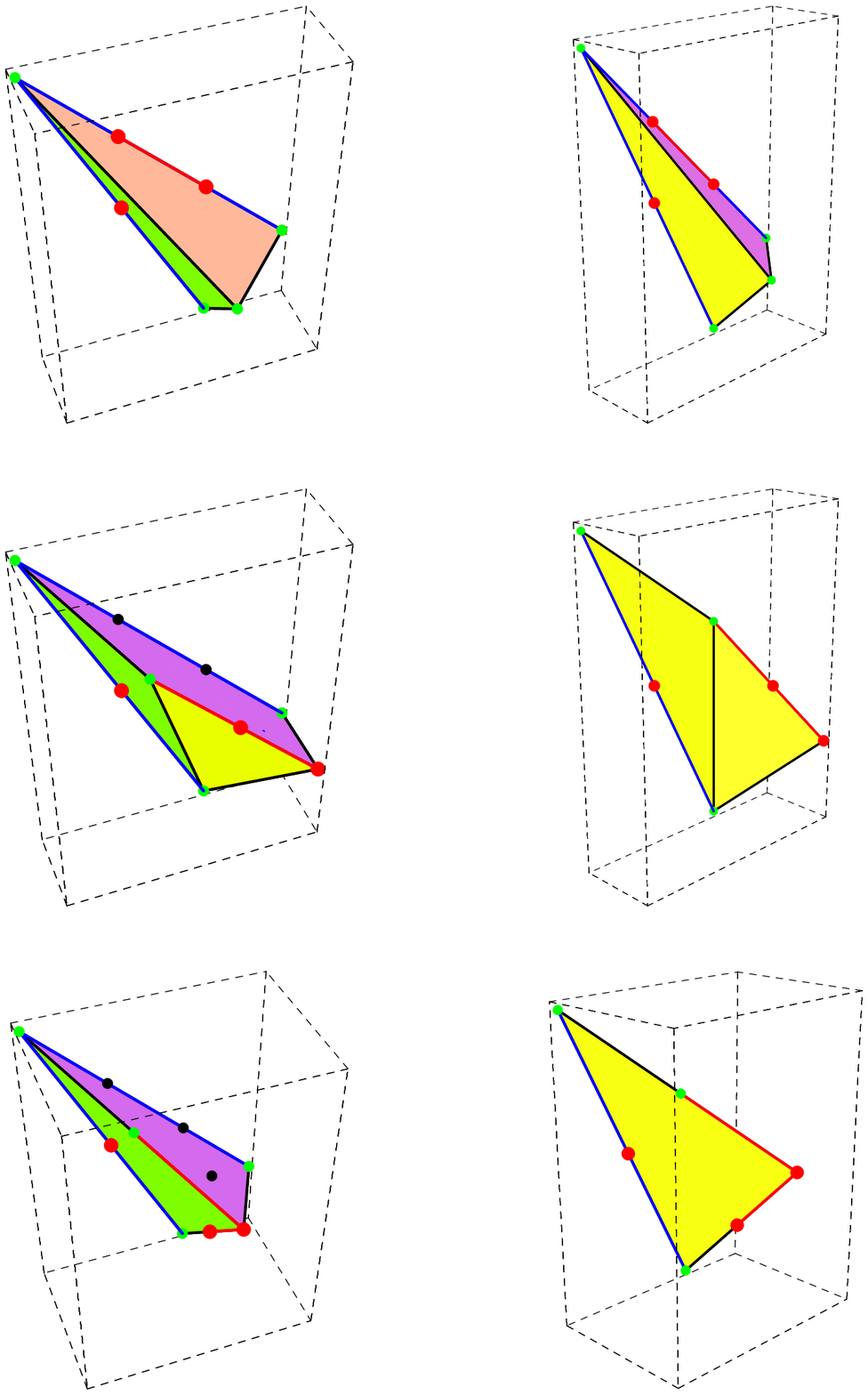}
\place{.3}{6.75}{$(SU_2{\times}SU_3)_c$}
\place{.3}{4.375}{$SO_5{\times}SU_2$}
\place{.3}{2}{$G_2{\times}SU_2$}
\newpage
\nonsimplefig{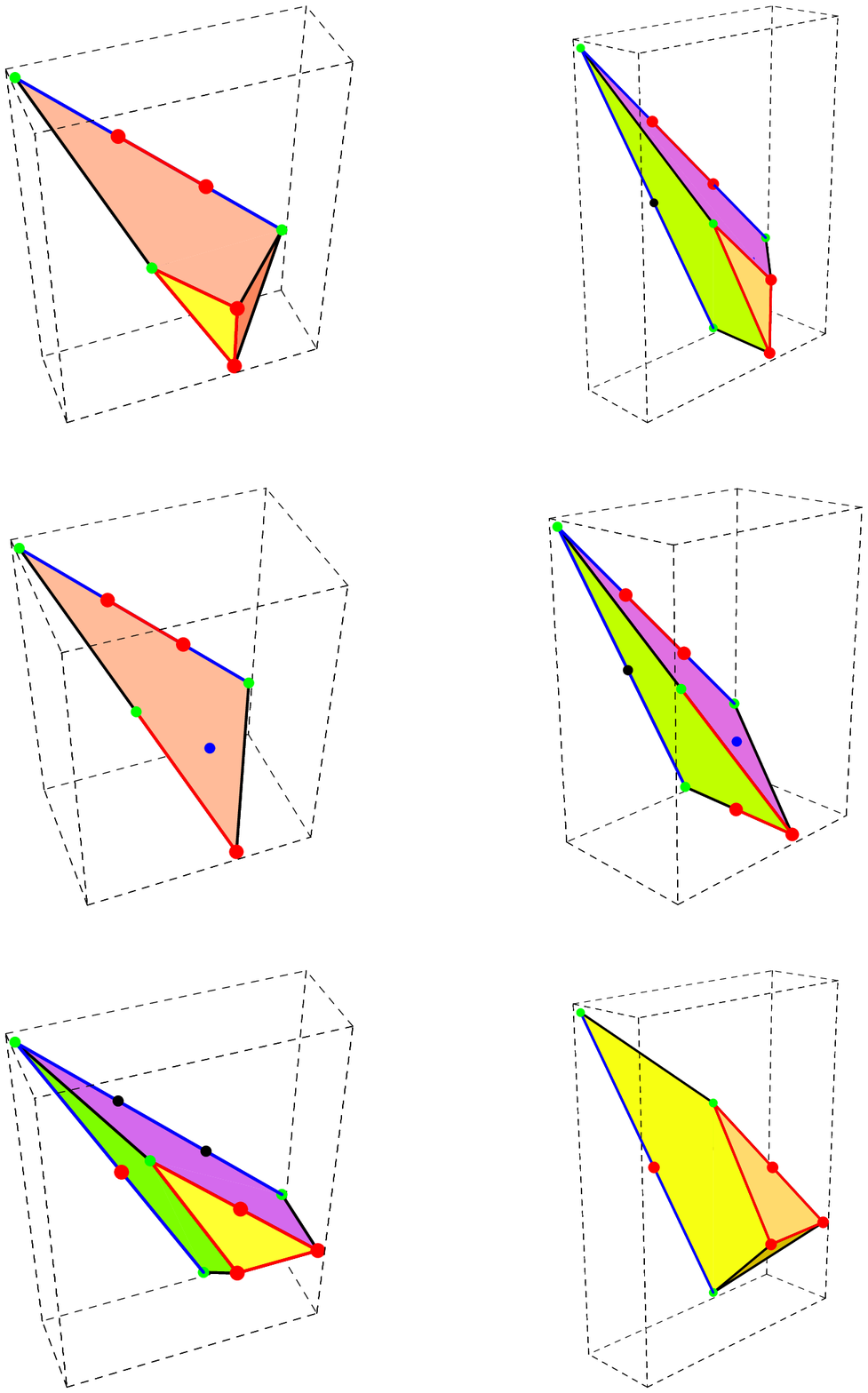}
\place{.3}{6.75}{$SU_3{\times}SU_3$}
\place{.3}{4.375}{$G_2{\times}SU_3$}
\place{.3}{2}{$SU_2{\times}SU_4$}
\newpage
\nonsimplefig{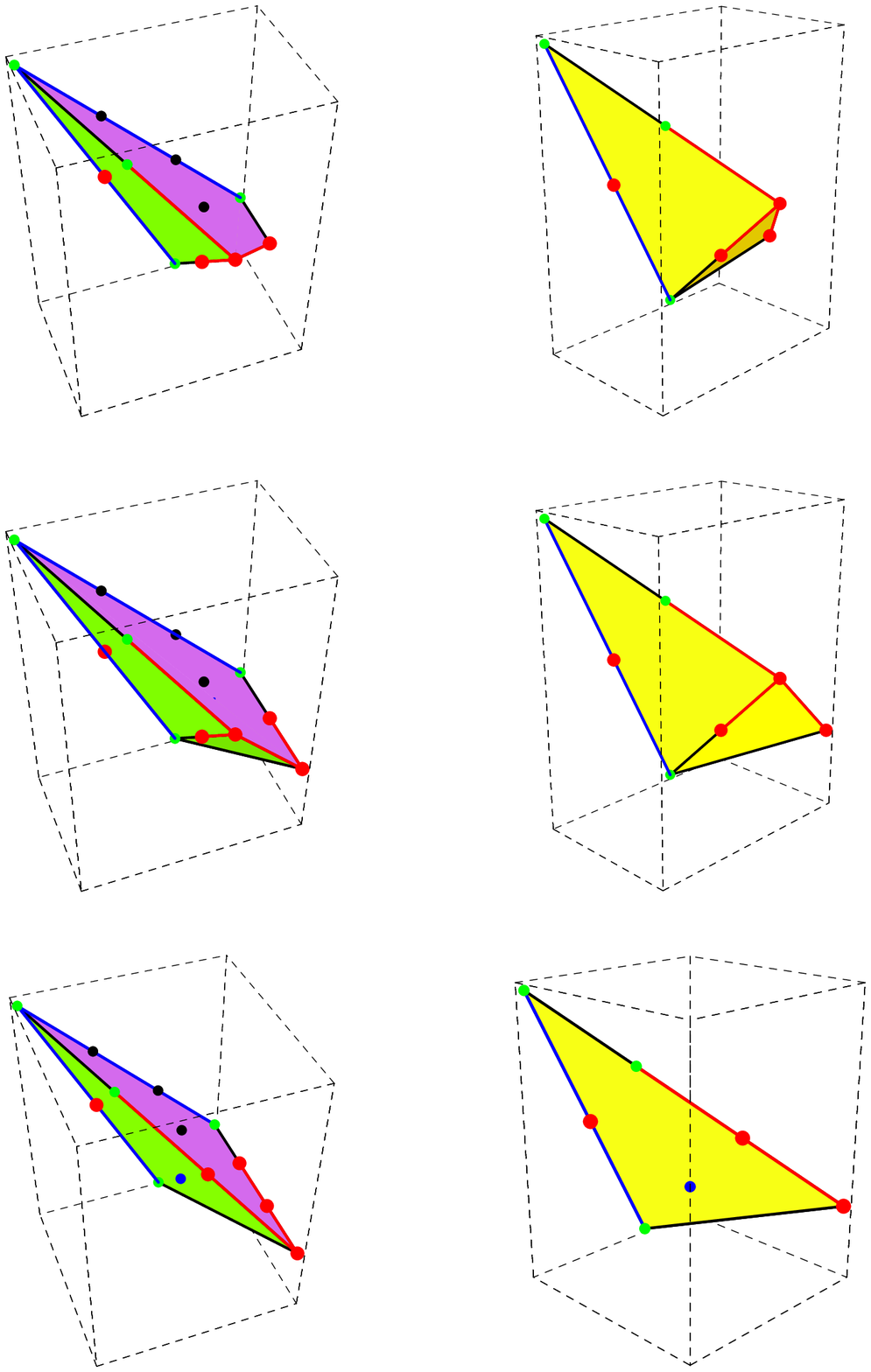}
\place{.3}{6.75}{$SO_7{\times}SU_2$}
\place{.3}{4.375}{$SO_9{\times}SU_2$}
\place{.3}{2}{$F_4{\times}SU_2$}
\newpage
\baselineskip=12.5pt plus 1pt minus 1pt
\immediate\closeout\referencewrite\referenceopenfalse
\line{\bf\hfil References\hfil}\bigskip\parindent=0pt\input referenc.texauxil

\bye